\documentclass[prd,aps,nofootinbib,preprint,eqsecnum,tightenlines]{revtex4}
\usepackage{graphicx,color,amsmath,amsxtra}
\usepackage{epsf}
\usepackage{amssymb}
\usepackage{enumerate}
\usepackage{hhline}
\usepackage{array}
\usepackage{tabularx}
\usepackage{booktabs}

\newcommand{\be}{\begin{equation}}
\newcommand{\ee}{\end{equation}}
\newcommand{\bea}{\begin{eqnarray}}
\newcommand{\eea}{\end{eqnarray}}
\newcommand{\beaa}{\begin{eqnarray*}}
\newcommand{\eeaa}{\end{eqnarray*}}

\def\be{\begin{equation}}
\def\ee{\end{equation}}
\def\bea{\begin{eqnarray}}
\def\eea{\end{eqnarray}}

\begin{document}

\title{Gravastars on the brane with a timelike extra dimension}

\author{Shounak Ghosh$^{1}$\footnote{E-mail address: shounakphysics@gmail.com}}
\author{Rikpratik Sengupta$^{2}$\footnote{E-mail address: rikpratik.sengupta@gmail.com}}
\author{Kazuharu Bamba$^{3}$\footnote{Corresponding author. E-mail address: bamba@sss.fukushima-u.ac.jp}}

\affiliation{
$^1$Directorate of Legal Metrology, Department of Consumer Affairs, Govt. of West Bengal, Dhupguri, Jalpaiguri 735210, West Bengal, India\\
$^2$Department of Physics, Indian Institute of Technology Kanpur, Kanpur 208016, Uttar Pradesh, India\\
$^3$Faculty of Symbiotic Systems Science, Fukushima University, Fukushima 960-1296, Japan
}


\begin{abstract}
We construct a gravastar configuration within the Shtanov-Sahni (SS) braneworld scenario with a timelike extra dimension and negative brane tension. The modified gravitational dynamics on the brane naturally regularize the interior geometry and prevent central curvature singularities. Solving the induced field equations, we obtain analytic interior, shell, and exterior solutions without invoking the thin-shell approximation. The gravastar core is modeled as a Bose--Einstein condensate, while the intermediate shell consists of ultra-dense stiff matter. Bulk Weyl corrections generate anisotropic effective pressures that arise intrinsically and support physical viability. We analyze the active gravitational mass, energy, entropy, proper shell thickness, and junction conditions at the interfaces. The configuration can exhibit suppressed or negative effective mass due to the repulsive interior condensate while remaining consistent with energy conditions, demonstrating that SS braneworld gravastars provide a viable compact-object alternative to classical black holes with singularity avoidance arising from higher-dimensional geometric effects.
\end{abstract}

\maketitle

\section{Introduction} \label{sec1}

The question of how gravitational collapse ultimately ends in astrophysics mirrors, in many respects, the problem of the initial state of the universe in cosmology. While Einstein’s General Relativity (GR) has enjoyed remarkable success in explaining a wide range of observational phenomena at intermediate energy scales in both astrophysical and cosmological contexts, its predictive power is severely challenged in regimes of extreme curvature and energy density. Such regimes naturally arise at the putative beginning of the universe and at the final stages of stellar gravitational collapse. In these limits, enormous amounts of energy are compressed into extremely small volumes, causing the energy density and curvature invariants to diverge. As a consequence, GR generically predicts spacetime singularities where the classical field equations lose their validity~\cite{Hawking1}. This pathological behavior strongly suggests that new physical effects, most plausibly of quantum origin, must become relevant in order to regularize the theory and remove these unphysical infinities.

In the context of stellar collapse, incorporating quantum field effects on a fixed classical spacetime background has already revealed several profound phenomena. The most celebrated example is Hawking radiation, whereby black holes emit thermal radiation from their event horizons~\cite{Hawking2}, often interpreted as the boundary separating the black hole interior from the external universe. Despite this remarkable insight, semiclassical gravity is insufficient to cure the singular behavior of classical black hole solutions, as the curvature singularity at the center persists unchanged~\cite{Birrell}. In the Schwarzschild solution, the apparent singularity at the Schwarzschild radius $R=2GM~(c=1)$ is merely a coordinate artifact, since all curvature invariants remain finite there and the singularity can be eliminated by an appropriate change of coordinates. In contrast, the divergence at $r=0$ corresponds to a genuine physical singularity, characterized by divergent curvature scalars, and cannot be removed by any coordinate transformation.

From a fundamental perspective, the emergence of physical singularities in the description of any physical system is widely regarded as signaling the breakdown or incompleteness of the underlying theoretical framework. Relativistic gravity presents two prominent and closely related examples of such singularities: those associated with cosmological models of the universe~\cite{Hawking} and those arising in black hole spacetimes~\cite{Penrose}. Both are curvature singularities, meaning that they are invariantly defined through divergences of curvature tensors and cannot be eliminated by coordinate redefinitions~\cite{HP}. In standard relativistic cosmology, a Ricci-type curvature singularity occurs at $t=0$, where the Riemann tensor and the energy density diverge, marking what is conventionally interpreted as the beginning of spacetime evolution~\cite{Hawking}. Analogously, classical black hole solutions contain a curvature singularity at $r=0$, where physical quantities blow up in an identical manner~\cite{Penrose}. This central singularity stands in sharp contrast to the horizon at $r=2M$, which represents only a removable coordinate singularity in Schwarzschild coordinates and can be regularized through alternative coordinate systems ~\cite{Finkelstein,Kruskal,Szekeres}.

In modern theoretical physics, it is increasingly accepted that the presence of such singularities reflects the limited domain of applicability of classical GR rather than a true physical feature of nature. This viewpoint is motivated by two key considerations. First, GR has not been experimentally tested in regimes of ultra-high curvature or energy density. Second, it is natural to expect that quantum effects should dominate the dynamics at such extreme scales, whereas GR is intrinsically a classical theory. Consequently, substantial effort has been devoted to extending or modifying GR in a manner that remains consistent with known observations while improving its high-energy behavior. One particularly influential strategy is to modify the geometric sector of the theory by generalizing the Einstein–Hilbert (EH) action. Originally introduced several decades ago as a means of incorporating ultraviolet (UV) corrections, such modifications have since been widely explored at both UV and infrared (IR) scales. This approach is often favored because it avoids many of the conceptual and phenomenological difficulties that arise when the matter sector alone is altered. Moreover, modifying the gravitational action may provide valuable clues toward the eventual construction of a self-consistent theory of quantum gravity (QG), even though such a complete theory remains elusive at present. In this regard, the early universe and compact objects like black holes constitute natural laboratories for probing the viability of QG-inspired models, as both are characterized by extreme curvature and energy density. Among the various approaches to quantum gravity, two broad and conceptually distinct frameworks have gained widespread attention. One of these is M-theory, which emerged as a unifying framework linking the five previously known superstring theories through duality relations and which reduces to supergravity in the low-energy limit~\cite{S2}. Quantum consistency of M-theory requires an eleven-dimensional spacetime, and its fundamental degrees of freedom include extended objects such as two- and five-dimensional membranes, commonly referred to as branes. Although M-theory has not yet been fully formulated, its structural features have inspired numerous developments in high-energy physics and cosmology.

Braneworld models have emerged as an important extension of higher-dimensional gravity, with the Randall--Sundrum (RS) \cite{Randall1,Randall2} playing a central role. This scenario provides an elegant explanation of the hierarchy between the gravitational and electroweak scales by postulating that our observable universe is a $(3+1)$-dimensional hypersurface embedded in a higher-dimensional spacetime. In this setup, Standard Model fields are confined to the brane, whereas gravity propagates in the surrounding five-dimensional bulk endowed with an Anti--de Sitter geometry ($AdS_5$). The presence of a negative bulk cosmological constant leads to the localization of the graviton near the brane. While the original RS proposal involved two branes, the later RS-II formulation \cite{Randall2} effectively removes the negative-tension brane by sending it to infinity, yielding a single-brane configuration. The Randall--Sundrum single--brane framework with a spacelike extra dimension has been widely investigated and applied to a variety of problems in cosmology~\cite{Binetruy,Maeda,Langlois,Chen,Kiritsis,Campos,Maartens} as well as in astrophysical contexts~\cite{Wiseman2,Germani,Wiseman,Visser,Creek,Pal,Bruni,Govender,Sengupta5}. Building on this line of research, Shtanov and Sahni proposed an alternative construction, commonly referred to as the SS model, which may be regarded as a dual of the RS--II scenario, where the spacelike extra dimension is replaced by a timelike one~\cite{Sahni4}. In the standard RS case the five--dimensional bulk possesses a Lorentzian signature, whereas introducing a timelike extra dimension leads to a bulk geometry with signature $(-,-,+,+,+)$. This change in the nature of the extra dimension significantly modifies the properties of the braneworld. Unlike the RS--II model, the SS setup is characterized by a negative brane tension and a positive bulk cosmological constant. One of its most remarkable features is that the cosmological evolution admits a smooth bounce during the contracting phase, thereby avoiding the formation of a singularity, even when the matter content on the brane satisfies the usual energy conditions. A potential concern in models with a timelike extra dimension is the emergence of tachyonic Kaluza--Klein gravitational modes, as originally discussed in Ref.~\cite{Sahni4}. This issue has, however, been subsequently examined and addressed in Ref.~\cite{S10}.

Against this broader backdrop, a novel alternative to classical black holes was proposed in the form of the gravitational vacuum condensate star, or gravastar~\cite{Mazur1}, which was further developed in subsequent work~\cite{Mazur2}. In parallel, related ideas suggested that quantum gravitational effects could transform the classical black hole horizon into a critical surface associated with a gravitational phase transition.
In this picture, the interior of the object is described by an equation of state (EoS) of the form $p=-\rho$~\cite{Gliner}, corresponding to a vacuum energy that exerts a repulsive gravitational effect capable of counteracting collapse. Building on these ideas, it was suggested that quantum fluctuations become dominant near the would-be horizon, particularly in the temporal and radial components of the energy–momentum tensor. These fluctuations can grow sufficiently large to effectively produce an EoS $p=\rho$, which represents the stiffest causal equation of state allowed by relativity. Such a state lies at the threshold of causality violation and naturally leads to the formation of a gravitational Bose–Einstein condensate (BEC) in the interior. The sharp horizon is then replaced by a thin shell composed of stiff matter, a concept first introduced by Zeldovich~\cite{Zeldo1,Zeldo2}. Outside this shell, the spacetime is assumed to be vacuum, with vanishing pressure and energy density. Early seminal works established primordial black holes (PBHs) as natural outcomes of scalar-field instabilities and first-order cosmological phase transitions, linking PBH formation to symmetry-breaking dynamics in the early universe \cite{Khlopov1985,Konoplich1999,Khlopov2000}. These studies demonstrated how bubble collisions, false-vacuum collapse, and density inhomogeneities can efficiently seed black holes across a wide mass spectrum. Subsequent reviews consolidated PBHs as probes of high-energy physics and cosmology beyond the Standard Model \cite{Khlopov2010}. More recently, regular black hole remnants with de Sitter cores and associated graviatoms were proposed as viable heavy dark matter candidates, sensitive to primordial inhomogeneities \cite{Dymnikova2015}.

The gravastar model thus emerges as a non-singular endpoint of gravitational collapse, designed specifically to eliminate the central curvature singularity that plagues classical black hole solutions~\cite{Mazur1}. In the simplest Schwarzschild black hole scenario, the event horizon is a null hypersurface across which physical quantities remain finite, allowing infalling observers to cross it smoothly. However, whether such analytic continuation remains valid once quantum effects are properly taken into account remains an open question. Motivated by this uncertainty, the classical event horizon was replaced by a thin shell separating distinct interior and exterior phases~\cite{Mazur2}. In the standard gravastar construction, the interior region is modeled as a vacuum condensate characterized by the equation of state ($p=-\rho$), which is mathematically equivalent to a cosmological constant and generates a repulsive gravitational effect.
Although this EOS violates the strong energy condition (SEC), $\rho+3p\geq0$, it still satisfies the null energy condition (NEC), $p+\rho\geq0$. Since violation of the NEC is a key ingredient in the Hawking–Penrose singularity theorems~\cite{HP}, it is not immediately obvious that this alone is sufficient to prevent singularity formation. Nevertheless, the violation of the SEC introduces repulsive effects that can significantly alter the collapse dynamics. To incorporate quantum considerations more fully, the horizon is treated as a locus of quantum phase transition, supported by an interior that violates the SEC and allows for a rearrangement of the vacuum state~\cite{Gliner}. The resulting configuration consists of a gravitational BEC core stabilized by an ultra-dense thin shell of stiff matter obeying $p=\rho$~\cite{Zeldo1,Zeldo2}. Comprehensive reviews of the gravastar paradigm can be found in Ref.~\cite{RS1}.

In recent years, gravastars have attracted considerable interest, and a variety of stable configurations have been constructed within both GR and modified gravity theories~\cite{G1,G2,G4,G5,G6,Sengupta5}. Such investigations are particularly relevant in the strong-field regime, where deviations from general relativity may become significant and potentially observable. The three-dimensional gravastar structures in massive gravity \cite{Barzegar:2023plb} and in gravity’s rainbow \cite{Barzegar:2023epjc} provide novel insights into lower-dimensional models, while extensions in $f(Q)$ gravity have been systematically explored through Chaplygin gas models \cite{Mohanty:2025ijmpd}, Krori–Barua metrics \cite{Mohanty:2024fdp}, and charged configurations \cite{Mohanty:2024aop}. Charged gravastars in $f(R,T)$ gravity have also been studied in detail \cite{Yousaf:2019prd}. A recent work explored gravastar configurations in $f(\mathcal{G},T)$ gravity, showing that Gauss–Bonnet curvature and matter–geometry couplings can support regular, stable horizonless compact objects with properties distinct from general relativity \cite{Shamir2018}. More recently, another work studied thin-shell gravastars in massive gravity using the Darmois–Israel formalism, demonstrating that a non-zero graviton mass significantly modifies shell stresses and physical viability conditions. Their results indicate that both higher-curvature corrections and massive gravity provide viable mechanisms for realizing stable, non-singular gravastar configurations without event horizons \cite{Das2024}.

Recent studies have significantly advanced the understanding of gravastars as viable horizonless compact objects across dynamical, perturbative, and observational contexts. It has been showed that gravastar formation via sudden vacuum condensation can be accompanied by thermal radiation with a characteristic de Sitter temperature, providing a possible radiative signature of gravastar birth \cite{Nakao2023}. Adler demonstrated that time-dependent gravastar configurations can evade several no-go results applicable to static exotic compact objects, underscoring the importance of dynamics for physical viability and light propagation \cite{Adler2024}. Some works have investigated observational imprints of thin-shell gravastars using ray-tracing simulations of accretion disks and hot spots, finding that gravastars can closely mimic black holes while still exhibiting potentially detectable deviations \cite{Rosa2024}. From a perturbative perspective, analyzed scalar quasinormal modes of gravastars have been analyzed in scalar–tensor theories, showing that additional scalar degrees of freedom modify the mode spectrum relative to general relativity \cite{Boumaza2025}. Earlier,it was established that gravastar oscillation spectra and gravitational-wave signatures differ qualitatively from those of black holes \cite{Pani2009}, while others have demonstrated that gravastars can approach the black-hole limit in their tidal deformability, clarifying how I–Love–Q relations for gravastars connect smoothly to those of black holes \cite{Uchikata2016}.

Since the avoidance of singularities is fundamentally a UV effect, it is particularly compelling to investigate gravastar solutions in frameworks that modify GR at high energies, where quantum gravitational corrections are expected to play a dominant role. The motivation for considering gravastars in a braneworld framework arises from the ultraviolet sensitivity of gravitational collapse, where classical four-dimensional general relativity is expected to receive significant corrections. While gravastar solutions can be constructed in GR, their physical viability typically requires imposed anisotropies, exotic equations of state, or infinitesimally thin shells. In braneworld gravity, the effective four-dimensional field equations acquire high-energy quadratic stress--energy corrections and non-local Weyl terms sourced by the bulk geometry, which become dominant at the energy scales relevant for compact objects. These geometrically induced contributions naturally generate effective repulsive gravity, intrinsic pressure anisotropy, and a reduced (or negative) active gravitational mass on the brane, thereby providing a theoretically grounded mechanism for stabilizing non-singular gravastars without introducing ad hoc matter sectors, while simultaneously offering potentially observable deviations from classical black hole phenomenology.

This work is devoted to the construction and analysis of a gravastar configuration within the framework of the SS braneworld scenario. The paper is structured as follows. In Sec.~\ref{sec2}, we present the effective field equations on a SS brane for a static, spherically symmetric spacetime and derive their explicit solutions by separately imposing appropriate equations of state for the interior region, the thin shell, and the exterior spacetime of the gravastar. A detailed examination of the resulting physical characteristics of the gravastar model is carried out in Sec.~\ref{sec3}. The matching conditions at the interfaces between different regions are established in Sec.~\ref{sec4}. Subsequently, the dynamical physical viability of the proposed gravastar configuration is analyzed in Sec.~\ref{sec5}. Finally, a summary of the main findings along with concluding remarks is provided in Sec.~\ref{sec6}.

\section{Mathematical formalism and solutions} \label{sec2}

A modification of the Einstein--Hilbert action in the geometric sector of braneworld gravity leads to the emergence of additional contributions in the effective field equations on the brane. The most general form of the modified Einstein field equations (EFE) induced on the brane can be expressed as~\cite{Sahni5}
\begin{equation} \label{eq1}
m^2 G_{\mu \nu} + \sigma h_{\mu \nu}
= T_{\mu \nu} + \epsilon M^3 \left( K_{\mu \nu} - h_{\mu \nu} K \right),
\end{equation}
where $m$ denotes the four-dimensional Planck mass, $M$ is the five-dimensional Planck mass, and $\sigma$ represents the brane tension.

Employing the Gauss--Codazzi relations on the brane, the above equation can be recast into the form~\cite{Sahni5}
\begin{equation}\label{eq2}
G_{\mu \nu} + \Lambda_{\mathrm{eff}} h_{\mu \nu}
= 8\pi G_{\mathrm{eff}} T_{\mu \nu}
+ \frac{\epsilon}{1+\beta}
\left[
\frac{S_{\mu \nu}}{M^6} - W_{\mu \nu}
\right].
\end{equation}

The dimensionless parameter $\beta$ is defined as
\begin{equation} \label{eq3}
\beta = \frac{2\sigma m^2}{3M^6}.
\end{equation}
Here, $G_{\mu \nu}$ is the Einstein tensor constructed from the induced metric on the $(3+1)$-dimensional brane, and $T_{\mu \nu}$ denotes the energy--momentum tensor of matter confined to the brane. The extrinsic curvature of the brane is denoted by $K_{\mu \nu}$, with trace $K = K^{\mu}_{\ \mu}$. The induced metric on the brane is given by
\begin{equation} \label{eq4}
h_{\mu \nu} = g_{\mu \nu} - \epsilon n_{\mu} n_{\nu},
\end{equation}
where $n^{\mu}$ is the inward-pointing unit normal vector to the brane and $g_{\mu \nu}$ is the bulk spacetime metric. The parameter $\epsilon = \pm 1$ specifies the nature of the extra dimension.

For $\epsilon = +1$, the bulk spacetime has Lorentzian signature and the extra dimension is spacelike, corresponding to the Randall--Sundrum type II (RS II) model. In contrast, choosing $\epsilon = -1$ renders the extra dimension timelike, resulting in the SS (SS) braneworld scenario. The effective cosmological constant and gravitational coupling on the brane are given by
\begin{equation}\label{eq5}
\Lambda_{\mathrm{eff}} = \frac{\Lambda_{\text{RS}}}{1+\beta},
\qquad
8\pi G_{\mathrm{eff}} = \frac{\beta}{m^2(1+\beta)},
\end{equation} 
where $\Lambda_{\text{RS}}$ denotes the cosmological constant on the RS II brane.

The terms enclosed within the square brackets in Eq.~(2.2) represent corrections arising due to extra-dimensional effects. The tensor $S_{\mu \nu}$ encodes quadratic contributions from the stress--energy tensor and is constructed from the bare Einstein equation
\begin{equation}\label{eq6}
E_{\mu \nu} = m^2 G_{\mu \nu} - T_{\mu \nu},
\end{equation}
derived from Eq.~(2.1). Explicitly, $S_{\mu \nu}$ takes the form
\begin{equation}\label{eq7}
S_{\mu \nu}
= \frac{1}{3} E E_{\mu \nu}
- E_{\mu \alpha} e^{\alpha}_{\nu}
+ \frac{1}{2}
\left(
E_{\alpha \gamma} e^{\alpha \gamma}
- \frac{1}{3} e^2
\right)
h_{\mu \nu},
\end{equation}
where $E = e^{\mu}_{\ \mu}$.

The remaining correction term $W_{\mu \nu}$ arises from the projection of the bulk Weyl tensor $W_{\mu \nu \alpha \gamma}$ onto the brane and captures the influence of bulk graviton degrees of freedom. This quantity is defined as
\begin{equation}\label{eq8}
W_{\mu \nu} = n^{\alpha} n^{\gamma} W_{\mu \nu \alpha \gamma}.
\end{equation}
Since $W_{\mu \nu}$ is traceless, it behaves as an effective radiation-like source on the brane, introducing additional effective energy density and pressure contributions in the modified field equations.

The correction terms satisfy the conservation condition
\begin{equation}\label{eq9}
D^{\mu}
\left(
S_{\mu \nu} - M^6 W_{\mu \nu}
\right) = 0,
\end{equation}
where $D^{\mu}$ denotes the covariant derivative compatible with the induced metric $h_{\mu \nu}$. This relation follows directly from the modified Einstein equations and need not be imposed as an independent constraint.

In the present work, we focus on the Randall--Sundrum limit, wherein the induced curvature term on the brane is absent. This is achieved by setting $m = 0$, reflecting the fact that quantum corrections to the matter action—which would otherwise generate an induced curvature term—are neglected in the RS framework. Although $m$ vanishes in this limit, gravity on the brane remains well-defined: the effective four-dimensional Planck scale is determined by a combination of bulk and brane parameters rather than by $m$ itself.

Conversely, taking the limit $M \to 0$ suppresses extra-dimensional contributions entirely, thereby recovering standard general relativity. It is noteworthy that for a timelike extra dimension, non-singular bouncing cosmological solutions can arise irrespective of whether the induced curvature term is present. When such a term is included, an additional class of solutions becomes possible in which the universe emerges from a quasi-singular state characterized by finite Hubble parameter and stress--energy tensor, while curvature invariants diverge.

In the RS limit ($m = 0$), the effective gravitational coupling on the brane is given by
\begin{equation}\label{eq10}
8\pi G_{\mathrm{eff}} = \frac{2\epsilon \sigma}{3M^6}.
\end{equation}
For the SS braneworld, where the extra dimension is timelike ($\epsilon = -1$), positivity of $G_{\mathrm{eff}}$ requires the brane tension $\sigma$ to be negative. The effective cosmological constant on the brane in this limit is
\begin{equation}\label{eq11}
\Lambda_{\text{RS}} = \frac{\Lambda_5}{2}
+ \frac{\epsilon \sigma^2}{3M^6},
\end{equation}
where $\Lambda_5$ denotes the bulk cosmological constant. A vanishing effective cosmological constant on the brane therefore necessitates a positive bulk cosmological constant when $\epsilon = -1$.

Following the original Randall-Sundrum construction, we impose the fine-tuning condition $\Lambda_{\text{RS}} = 0$ in order to develop our gravastar model on the brane. Since the extra dimension is timelike, the evolution off the brane into the bulk constitutes a well-posed Cauchy problem with a valid solution in a neighborhood of the brane. Consequently, one may adopt an effective approach that focuses exclusively on the brane dynamics without specifying the global bulk geometry~\cite{S9}.

The conservation law governing the matter sector on the brane remains unaltered, and from Eq. (2.9) it can be expressed in the standard form
\begin{equation} \label{eq12}
\frac{dp}{dr}
= -\frac{1}{2}\frac{d\nu}{dr}(\rho + p).
\end{equation}

In order to construct gravastar configurations, it is necessary to solve the modified Einstein field equations on the brane assuming a static and spherically symmetric spacetime geometry. The corresponding line element is chosen as
\begin{equation} \label{eq13}
ds^2
= -e^{\nu(r)}dt^2
+ e^{\lambda(r)}dr^2
+ r^2 \left( d\theta^2 + \sin^2\theta  d\phi^2 \right).
\end{equation}

It follows from Eq.~(\ref{eq1}) that
\begin{eqnarray}
&&e^{-\lambda}\left(\frac{\lambda'}{r}-\frac{1}{r^2}\right)+\frac{1}{r^2} =\rho^\mathrm{eff},\label{eq14}\\ 
&&e^{-\lambda}\left(\frac{\nu'}{r}+\frac{1}{r^2}\right) -\frac{1}{r^2} =p_r^\mathrm{eff},\label{eq15}\\
&&e^{-\lambda}\left(\frac{\nu''}{2}-\frac{\lambda' \nu'}{4}+\frac{\nu'^2}{4}+\frac{\nu'-\lambda'}{2r}\right) = p_t^\mathrm{eff}.\label{eq16}
\end{eqnarray}

The effective density and effective radial and tangential pressures can be expressed as
\begin{eqnarray}\nonumber
&&\rho^\mathrm{eff}=8
\pi \rho \left( r \right)  \left[ 1-{\frac {\rho \left( r \right)
}{ \rho_\mathrm{c}}} \right] -{\frac {12U}{ \rho_\mathrm{c}}},\\ \nonumber
&&p_r^\mathrm{eff}=8
\pi  \left[ p \left( r \right) -{{\frac {\rho \left( r \right)
}{ \rho_\mathrm{c}}} \left(2 p \left( r \right) +\rho \left( r \right)  \right)  } \right] -{\frac {4U}{ \rho_\mathrm{c}}}-{\frac {8P}{ \rho_\mathrm{c}}},\\ \nonumber
&&p_t^\mathrm{eff}=8\pi  \left[ \mu\rho \left( r
 \right) -{\frac {\rho \left( r \right)  \left( 2\mu\rho \left( r
 \right) +\rho \left( r \right)  \right) }{ \rho_\mathrm{c}}} \right] -
{\frac {4U}{ \rho_\mathrm{c}}}+{\frac {4P}{ \rho_\mathrm{c}}}.\nonumber
\end{eqnarray}

The parameter $\rho = \rho_\mathrm{c}$ is introduced as a constant critical energy density that characterizes the occurrence of a cosmological bounce. In the present framework, this critical density is related to the brane tension through
\begin{equation} \label{eq17}
\rho_\mathrm{c} = 2|\sigma|,
\end{equation}
where $\sigma$ denotes the brane tension. The absolute value is required since the SS (SS) braneworld scenario involves a negative brane tension.

The quantities $U$ and $P$ appearing in the modified Einstein field equations represent effective stress--energy components arising from the projection of the bulk Weyl tensor onto the brane. Specifically, $U$ corresponds to the effective energy density, while $P$ denotes the effective pressure induced on the brane due to bulk gravitational effects. A noteworthy feature of this construction is that, despite assuming ordinary matter on the brane to be a perfect fluid with isotropic pressure, the higher-dimensional contributions generate unequal effective radial and tangential pressures. This leads to an induced anisotropy in the effective pressure on the brane, quantified by a difference of magnitude $\frac{12P}{\rho_\mathrm{c}}$, entirely originating from bulk effects. This observation supports the viewpoint that pressure anisotropy is a fundamental characteristic of gravastar configurations~\cite{Cattoen}. While such anisotropy does not arise naturally within the standard relativistic description of perfect fluids, it emerges here as an inherent consequence of the underlying framework.

Furthermore, the mean effective pressure associated with the projected Weyl contribution can be written as
\begin{equation}\label{eq18}
\tilde{P}_{\mathrm{eff}}^{\text{avg}}
= \frac{1}{3}
\left(
\tilde{P}^{r}_{\mathrm{eff}} + 2 \tilde{P}^{t}_{\mathrm{eff}}
\right)
= \frac{1}{3} \tilde{\rho}_{\mathrm{eff}},
\end{equation}
which clearly demonstrates that the additional effective matter content induced by the Weyl projection obeys a radiation-like equation of state. This behaviour is a direct consequence of the traceless nature of the projected Weyl tensor.

The matter content confined to the brane is modeled as a perfect fluid, whose stress--energy tensor is given by
\begin{equation}\label{eq19}
T^{\mu}_{\nu} = \text{diag}(-\rho, p, p, p).
\end{equation}
The energy density $\rho$ and isotropic pressure $p$ are related via a linear equation of state parameterized by $\mu$, such that
\begin{equation} \label{eq20}
p(r) = \mu  \rho(r).
\end{equation}
The value of the parameter $\mu$ is determined by imposing the Israel--Darmois junction conditions at the shell of the gravastar which acts a mathematical matching hypersurface or junction.

As discussed earlier, we adopt an effective brane-based approach~\cite{S9} in which the detailed dynamics of the bulk spacetime are not explicitly specified. This allows considerable freedom in selecting the functional form of the projected Weyl quantities on the brane. Similar assumptions regarding the Weyl sector have been employed ~\cite{Sengupta5,Sengupta2}. In the present analysis, we assume a linear equation of state relating the Weyl-induced pressure and energy density of the form
\begin{equation}\label{eq21}
P = w U,
\end{equation}
where $w$ is a constant parameter to be fixed using the junction conditions.

Finally, the energy density associated with the projected Weyl tensor on the brane is taken to follow the profile
\begin{equation}\label{eq22}
U(r) = \sqrt{\rho_0  \rho(r)},
\end{equation}
with $\rho_0$ being an additional constant model parameter. Its value is determined through the boundary conditions imposed at the gravastar surface.

For the spacetime geometry, the temporal metric component is assumed to follow the regular Kuchowicz ansatz~\cite{Kuchowicz}, given by
\begin{equation}\label{eq23}
e^{\nu(r)} = \exp({B r^{2} + 2 \ln C}),
\end{equation}
where $B$ and $C$ are constant parameters. The constant $B$ carries dimensions of inverse length squared, whereas $C$ is a dimensionless quantity. The Kuchowicz potential is adopted primarily due to its regular and well-behaved nature over the entire range of finite radial coordinates. This property makes it particularly suitable for modeling the interior metric potential of non-singular compact configurations, such as gravastars and other regular collapse solutions~\cite{Biswas2020}. Furthermore, the versatility of the Kuchowicz potential has been demonstrated in other gravitational settings, where it has successfully been employed as a redshift function in the construction of wormhole geometries~\cite{Sengupta2}.

\subsection{Interior solution}

As discussed earlier, the gravastar core is characterized by an equation of state of the form
\begin{equation}
p = -\rho .
\end{equation}
This equation of state effectively mimics a dark-energy–like fluid and plays a crucial role in halting further gravitational collapse. Following the phase transition at the would-be event horizon—replaced in the gravastar picture by a thin shell—the interior is assumed to settle into a gravitational Bose–Einstein condensate. The negative pressure associated with this condensate generates an outwardly directed repulsive force, counterbalancing the inward pull of gravity and thereby stabilizing the configuration.

Substituting this equation of state into the conservation equation immediately yields
\begin{equation}
p = -\rho = -\rho_\mathrm{c} ,
\end{equation}
where $\rho_\mathrm{c}$ denotes the constant energy density of the gravastar interior. Consequently, the pressure is also spatially uniform throughout the core.

To determine the interior spacetime geometry, these constant values of pressure and energy density must be inserted into the right-hand side of the modified gravitational field equations on the Shtanov–Sahni brane. In doing so, the matter sector of the effective brane equations is fully specified by the interior gravastar fluid, allowing the metric potentials to be computed consistently within this framework.

For the gravastar interior, the inverse of the unknown metric potential corresponding to the radial coordinate turns out to have the form

\begin{equation}
\label{eq25}
e^{-\lambda(r)} =
\frac{
8\pi C_1^{2} r^{2}
- 8\pi \rho_{\mathrm{c}} C_1 r^{2}
- 4 r^{2} (2w + 1)\sqrt{\rho_{0} C_1}
+ \rho_{\mathrm{c}}
}{
\rho_{\mathrm{c}}\left(2 B r^{2} + 1\right)
}
\end{equation}

As we see, it is found to depend on the critical energy density $\rho_\mathrm{c}$ which puts an upper bound on the energy density contained in the brane, in terms of the fundamental physical parameter brane tension $\sigma$. In the high energy limit, the constant brane tension tends to acquire very small values, such that extra dimensional effects are significant. In the low energy limit $\sigma \rightarrow \infty$ and standard GR is recovered. Besides the brane tension, the metric potential also depends on the EoS parameter $w$, which basically describes an effective fictitious fluid on the brane, that basically models the non-local gravitational corrections arising from transfer of gravitational degrees of freedom from the bulk to the brane. There is also an additional dependence on the constant of integration $C_1$.   

\subsection{Active gravitational mass $M(R)$}

The active gravitational mass has the form
	\begin{align}
\label{eq26}
M(R_1)
&= \int_{0}^{R_1} 4\pi r^{2}\rho_{\mathrm{eff}}dr \nonumber \\
&= \frac{32}{3}\pi^{2} C_1 R_1^{3}
\left(1 - \frac{C_1}{\rho_{\mathrm{c}}}\right)
- \frac{12}{\rho_{\mathrm{c}}}\sqrt{\rho_{0} C_1}
\end{align}

This definition generalizes the standard Tolman–Whittaker notion of active mass to the effective four-dimensional description on the Shtanov–Sahni brane, where $\rho_{\mathrm{eff}}$ includes both the matter contribution and high-energy corrections.

For the gravastar core, the equation of state $p=-\rho=-\rho_\mathrm{c}$ implies a vacuum-like interior with constant energy density and negative pressure. In general relativity, the active gravitational mass density is governed by the combination $\rho+3p$, which for this equation of state becomes
\begin{equation}
\rho+3p=-2\rho_\mathrm{c}<0 ,
\end{equation}
signaling a repulsive gravitational behavior. This property is inherited by the effective description on the brane and is encoded in the structure of $\rho_{\mathrm{eff}}$.

The explicit result reflects this repulsive character. The leading contribution, proportional to $R_1^3$, resembles the usual volume term of a constant-density configuration but is modified by the factor $\left(1-C_1/\rho_\mathrm{c}\right)$, which suppresses the effective mass due to the negative pressure of the interior condensate. The second term enters with a negative sign and further reduces the total mass, arising from the high-energy brane corrections encapsulated in $\rho_{\mathrm{eff}}$.

Consequently, the total active gravitational mass of the gravastar interior can be strongly suppressed and may even become negative for suitable choices of the parameters $C_1$ and $\rho_\mathrm{c}$. This negative or vanishing active mass provides a concrete dynamical realization of the repulsive nature of the gravastar core, preventing continued collapse and ensuring the absence of a central singularity. It is this feature that fundamentally distinguishes the gravastar interior from that of a classical black hole, even though both may share the same exterior geometry.

\subsection{Intermediate Shell}

To obtain the solution of the metric function $g_{rr}$ for the shell, we make use of the EoS $p=\rho$, yielding 

\begin{equation}\label{eq27}
e^{-\lambda(r)} =
\frac{-24\pi C_1^{2} r^{2}e^{-2Br^{2}}
 +{8\pi C_1 r^{2}\rho_{\mathrm{c}}e^{-Br^{2}}}
- 4 r^{2}(2w+1)\sqrt{\rho_{0} C_1}
e^{-\frac{Br^{2}}{2}}
+ \rho_{\mathrm{c}}}{\rho_{\mathrm{c}}\left(2Br^{2}+1\right)}.
\end{equation}

It becomes necessary for considering that the shell is composed of extremely dense stiff matter so that it can support the repulsive effect of the interior described by the gravitational BEC and still maintain a stable configuration. However, considering the matter constituting the shell to be stiff matter, it often becomes necessary in almost all modified gravity as well as relativistic scenarios to assume the thin shell approximation. The region separating the interior and exterior geometries of the gravastar is modeled as a junction layer connecting the two spacetimes. Although this layer has a finite thickness, it is assumed to be extremely narrow compared to the characteristic length scales of the system. In this limit, $e^{-\lambda}\ll 1$. However, it may be noted that in our present gravastar construction, analytic solutions of the modified gravitational field equations in the shell region can be obtained without invoking the idealized thin-shell approximation. As a consequence, the junction layer need not be treated as infinitesimally thin. Instead, the model naturally accommodates a shell with finite thickness, offering a more realistic and physically motivated description of the transition between the interior and exterior spacetimes. This feature enhances the internal consistency of the model and improves its physical plausibility compared to approaches that rely on an idealized, zero-thickness shell. 

\subsection{Exterior solution}

The EoS for the exterior region of the gravastar is $p=\rho=0$. The exterior region of the gravastar corresponds to a vacuum spacetime. However, unlike standard general relativity, the vacuum on the brane is not entirely empty once the influence of the higher-dimensional bulk is taken into account. In particular, the projection of the bulk Weyl tensor onto the brane gives rise to additional effective stress–energy components, which modify the exterior geometry through the appearance of a tidal charge.

As a result, the exterior line element is no longer purely Schwarzschild but takes the modified form
\begin{equation}
ds^2 = -\left(1-\frac{2M}{r}-\frac{Q}{r^2}\right)dt^2
+ \left(1-\frac{2M}{r}-\frac{Q}{r^2}\right)^{-1}dr^2
+ r^2\left(d\theta^2+\sin^2\theta d\phi^2\right),
\end{equation}
where $M$ denotes the gravitational mass measured by an exterior observer, while $Q$ represents the tidal charge induced by bulk gravitational effects. The parameter $Q$ is dimensionless and originates entirely from the effective stress–energy contributions associated with the projected Weyl tensor on the brane.

Although the exterior metric employed in the present work is mathematically similar to the Reissner--Nordström geometry, the physical interpretation of the parameter $Q$ is fundamentally different from that of the electric charge appearing in electrically charged gravastar models. In charged gravastar configurations constructed within General Relativity, the Reissner--Nordström exterior arises because the spacetime is sourced by a Maxwell electromagnetic field, and the charge parameter is directly associated with a conserved electric charge. By contrast, the quantity $Q$ appearing in Eq.~(2.30) is a \emph{tidal charge} generated by the projection of the five-dimensional Weyl tensor onto the four-dimensional brane. It therefore originates from nonlocal gravitational effects associated with the bulk geometry rather than from any electromagnetic source. From the effective four-dimensional perspective, the projected Weyl tensor acts as an additional gravitational source that carries information about the higher-dimensional bulk spacetime. Consequently, the term proportional to $Q/r^2$ may mimic the mathematical structure of the Reissner--Nordström metric while possessing an entirely different physical origin. In this sense, the exterior geometry can be viewed as an effective Reissner--Nordström-type spacetime generated purely by gravitational effects arising from the extra dimension. An important distinction is that, unlike the square of the electric charge in the Reissner--Nordström solution, the tidal charge parameter is not constrained by electromagnetic considerations and may in principle take either sign. Its value is determined by the bulk gravitational field and the embedding of the brane in the higher-dimensional spacetime. Thus, while electrically charged gravastars and braneworld gravastars may share certain mathematical similarities at the level of the exterior metric, the underlying physics is fundamentally different. The tidal charge parameter should be interpreted as a manifestation of nonlocal bulk gravity rather than an electromagnetic charge. This distinction highlights one of the principal differences between the present braneworld gravastar model and charged gravastar configurations previously studied in the literature.

The influence of the bulk gravity is therefore not confined to the interior of the gravastar but extends into the exterior vacuum region as well. If the effective correction to the exterior energy density arising from the Weyl term is denoted by $U_{\text{e}}$, it scales inversely with the fourth power of the radial coordinate, such that $U_{\text{e}} \sim Q/r^4$. Consequently, the sign of the tidal charge is directly determined by the sign of $U_{\text{e}}$: a positive effective energy density corresponds to a positive tidal charge, while a negative $U_{\text{e}}$ leads to a negative value of $Q$.

In conventional braneworld models with a spacelike extra dimension, the effective energy density $U_{\text{e}}$ is typically negative in order to ensure the localization of gravity on the brane, which in turn implies a negative tidal charge. In contrast, the present analysis is carried out within a braneworld framework featuring a timelike extra dimension. Consistent with this choice, the positive branch of the effective energy density is adopted, both in the interior and in the exterior regions. The appearance of the negative sign in front of the tidal charge term in the metric is itself a direct consequence of the timelike nature of the extra dimension.

Even when the tidal charge assumes only a small nonzero value, whether positive or negative, it introduces significant ultraviolet corrections to the predictions of standard general relativity. Within the Shtanov–Sahni braneworld scenario, the presence of a negative brane tension allows a positive effective energy density to confine gravity to the brane, thereby favoring a positive tidal charge. Had the effective energy density been negative, as in the Randall–Sundrum framework, the resulting exterior geometry would have closely resembled that of the Reissner–Nordström spacetime.

\section{Physical properties of the gravastar} \label{sec3}

A thorough examination of the physical characteristics of the gravastar shell is a crucial step in assessing the viability of the model. Of particular importance are the effective energy density and the associated pressures in both the radial and tangential directions of the matter constituting the shell, as well as global quantities such as the total energy, entropy, and the proper thickness of the layer.
In what follows, we investigate these properties for the gravastar shell, which acts as a transition region connecting the interior core to the exterior vacuum geometry. Since the shell possesses a finite thickness, all physical quantities defined within it must remain finite, regular, and smoothly varying across the junction. Ensuring the continuity and well-behaved nature of these quantities throughout the shell is essential for establishing a self-consistent and physically acceptable gravastar configuration.

\subsection{Matter density and pressure of the shell}

Employing the equation of state for the shell material, the energy density within the shell can be written as

\begin{equation}\label{eq28}
\rho = C_1 e^{-\nu} = C_1 \exp\left[-\frac{1}{2}\frac{(\mu+1) B r^2}{\mu}\right].
\end{equation}

It then follows that, upon enforcing the equation of state $p=\rho$, the effective matter density can be explicitly evaluated as

\begin{equation}\label{eq29}
\rho_\mathrm{eff}=8\pi \rho \left( r \right)  \left[ 1-{\frac {\rho \left( r\right) }{ \rho_\mathrm{c}}} \right] -{\frac {12U}{ \rho_\mathrm{c}}}.
\end{equation}

We compute the effective radial pressure  and it turns out to have the form

\begin{equation}\label{eq30}
p_r^\mathrm{eff}=\pi  \left[ p \left( r \right) -{\frac {\rho \left( r \right)
 \left(2 p \left( r \right) +\rho \left( r \right)  \right) }{\rho
_{{c}}}} \right] -{\frac {4U}{ \rho_\mathrm{c}}}-{\frac {8P}{ \rho_\mathrm{c}}}.
\end{equation}

The effective tangential pressure is also computed and can be expressed as

\begin{equation}\label{eq31}
p_t^\mathrm{eff}=8\pi  \left[ \mu\rho \left( r \right) -{\frac {\rho \left( r
 \right)  \left( 2\mu\rho \left( r \right) +\rho \left( r
 \right)  \right) }{ \rho_\mathrm{c}}} \right] -{\frac {4U}{ \rho_\mathrm{c}}}+
{\frac {4P}{ \rho_\mathrm{c}}}.
\end{equation}

\begin{figure}[t!]
\centering 
\includegraphics[width=7cm]{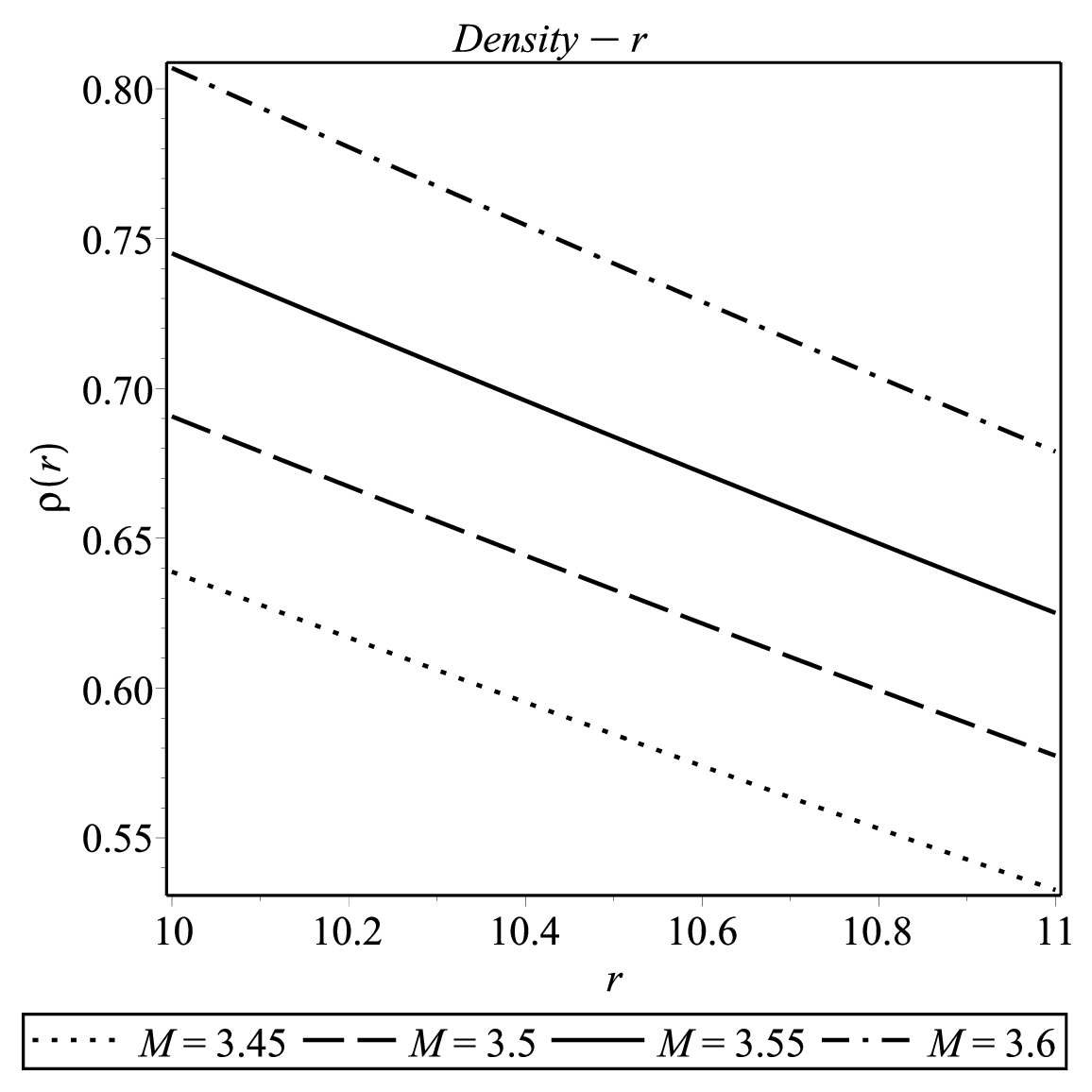}
\caption{Variation of the effective energy density of the gravastar shell, $\rho_{\rm eff}$, as a function of the radial coordinate $r$ within the finite shell region $R\leq r\leq R+\epsilon$. The different curves represent the profile of $\rho_{\rm eff}(r)$ obtained from Eq.~(3.2) for different gravastar masses. The numerical evaluation is performed using the fixed parameters $Q=0.138221\,\mathrm{km}^{2}$ and $\rho_0=0.05\times10^{-7}\,\mathrm{km}^{-6}$ together with the model parameters listed in Table~I. The plot shows that the effective energy density attains its maximum value at the inner boundary of the shell and decreases monotonically toward the outer boundary, indicating a smooth and regular matter distribution across the shell.}\label{pres.} 
\end{figure} 

The variation of the effective energy density with the radial distance $r$ along the shell of the gravastar is plotted in Fig.~1. As we find, the energy density is maximum at the shell-interior interface and keeps on decreasing monotonically in a linear way as we proceed towards the exterior boundary of the shell. 

\subsection{Energy}

The energy associated with the shell is obtained by performing a radial integration of the effective matter density over the thickness of the shell, which leads to

\begin{eqnarray}\label{eq32}
E &=& 4\pi \int_{R}^{R+\epsilon}\rho^\mathrm{eff}r^2 dr  \nonumber\\
&=& 4\pi {C_1} \left[ -{\frac {r\mu}{ \left( \mu+1 \right) B}{
{e}^{-{\frac { \left( \mu+1 \right) B{r}^{2}}{2\mu}}}}}+{
\frac {\mu\sqrt {2\pi }}{ 2\left( \mu+1 \right) B}{\text{erf}}
\left(\sqrt {{\frac { \left( \mu+1 \right) Br}{2\mu}}}
\right){\frac {1}{\sqrt {{\frac { \left( \mu+1 \right) B}{\mu}}}}}}\right],
\end{eqnarray}
where $\text{erf}$ denotes the error function.

\begin{figure*}[t!]
\centering
\includegraphics[width=7cm]{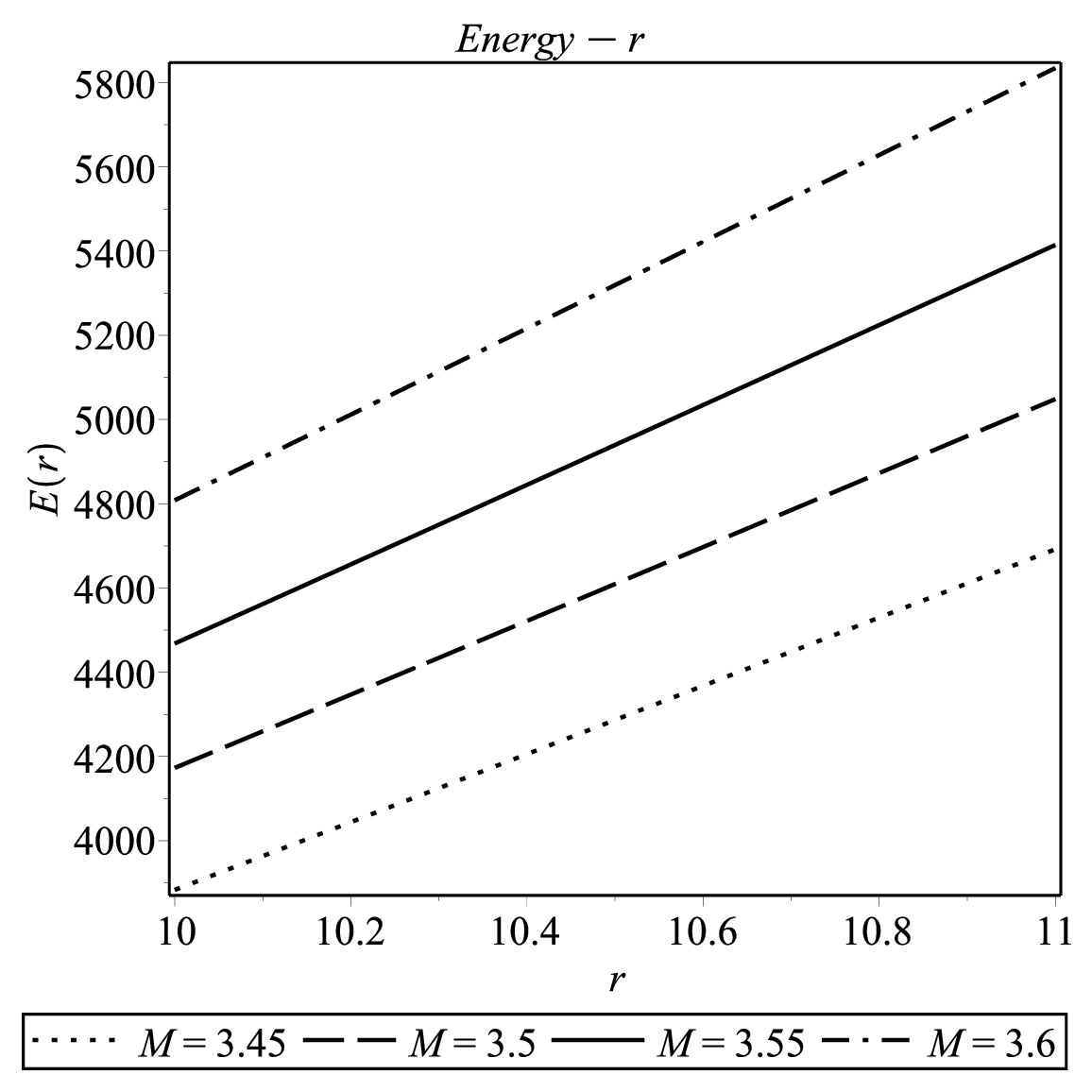}
\caption{Energy contained within the gravastar shell, $E$, as a function of the radial coordinate $r$ obtained from Eq.~(3.6). The different curves correspond to different gravastar masses and are computed using the fixed parameters $Q=0.138221\,\mathrm{km}^{2}$ and $\rho_0=0.05\times10^{-7}\,\mathrm{km}^{-6}$ together with the model parameters listed in Table~I. The energy increases monotonically with increasing $r$ from the interior boundary $r=R$ to the exterior boundary $r=R+\epsilon$, reflecting the gradual accumulation of effective matter energy throughout the finite shell region.}
\end{figure*}

With $p=\rho$, we have

\begin{eqnarray}\label{eq33}
E=4\pi {C_1} \left[ -{\frac {r{{e}^{-B{r}^{2}}}}{2B}}
+{\frac {\sqrt {\pi }{\text{erf}} \left(\sqrt {B}r\right)}{4{B}^{3/2}
}} \right].
\end{eqnarray}

As we can see in Fig.~2, the energy of the shell increases monotonically in a linear manner as we move from the interior to the exterior boundary of the gravastar. Physically, this indicates a stable and well-ordered transfer of gravitational stress from the repulsive interior to the exterior vacuum, ensuring consistent matching at both boundaries. Such behavior supports the interpretation of the shell as a realistic finite-thickness layer capable of maintaining equilibrium and preventing collapse or inphysical viability.

\subsection{Entropy}

Entropy is a fundamental quantity in black hole thermodynamics, and a corresponding analysis is therefore essential for a gravastar configuration. Accordingly, we evaluate the entropy of the gravastar on the brane using the relation
\begin{equation}\label{eq48}
S=\int_R^{R+\epsilon} 4 \pi r^2 s(r) \sqrt{e^{\lambda}} dr,
\end{equation}
where the entropy density $s(r)$ is given by
\begin{equation}\label{eq49}
 s(r)=\frac{\xi^2 k_B^2 T(r)}{4\pi\hbar^2}=\frac{\xi k_B}{\hbar}\sqrt{\frac{p}{2\pi}},
\end{equation}
 $\xi$ being constant having no dimension. In natural units, this density reduces to
 \begin{equation}\label{eq50}
 s(r)=\xi\sqrt{\frac{p}{2\pi}}.
\end{equation}

The shell entropy for the  gravastar has the form
\begin{align}\label{eq51}
S
&= \frac{4 \pi r^{2} \alpha k_{B} \epsilon}{\hbar}
\left[
\frac{2 B r^{2} \rho_{\mathrm{c}} + \rho_{\mathrm{c}}}{
-24 r^{2} C_1^{2} \pi e^{-2 B r^{2}}
+ 8 \pi C_1 r^{2} \rho_{\mathrm{c}} e^{-B r^{2}}
- 8 r^{2} \left( w + \frac{1}{2} \right)
\sqrt{\rho_{0} C_1} e^{-\frac{1}{2} B r^{2}}
+ \rho_{\mathrm{c}}
}
\right]^{1/2}.
\end{align}

The shell entropy of the gravastar is found to depend on the brane tension and effective EOS parameter mimicking bulk gravitational contributions on the brane. This reflects the fact that in our model the microscopic degrees of freedom of the gravastar are not purely four-dimensional.The brane tension controls the strength of high-energy corrections to the effective field equations on the brane. A lower (or negative) brane tension enhances the influence of bulk gravity, increasing the effective energy density and pressure within the shell. This, in turn, raises the number of accessible microstates of the shell matter and leads to a larger entropy. In the limit of very large brane tension, these corrections are suppressed, and the shell entropy approaches its standard general-relativistic behavior.

The effective equation-of-state parameter encodes how bulk gravitational effects, transmitted through the projected Weyl tensor, modify the local thermodynamic properties of the shell matter. By mimicking an additional matter component induced by the bulk, this parameter alters the balance between pressure and energy density inside the shell, directly affecting the entropy density and its radial distribution. Together, these dependencies highlight that the gravastar entropy is not purely geometric, as in the Bekenstein–Hawking description of black holes, but instead arises from the physical degrees of freedom of the finite-thickness shell. The shell entropy therefore provides a clear thermodynamic imprint of extra-dimensional gravity and distinguishes the gravastar from a classical black hole both dynamically and thermodynamically.

\subsection{Proper thickness}

The proper radial thickness of the shell, accounting for the curved geometry, can be expressed as

\begin{align}\label{eq38}
\ell
&= \int_{R}^{R+\epsilon} \sqrt{e^{\lambda}} dr
 = \epsilon \sqrt{e^{\lambda}} \nonumber\\[1ex]
&= \epsilon \Bigg[
\frac{
2 B r^{2} \rho_{\mathrm{c}} + \rho_{\mathrm{c}}
}{
\begin{aligned}
&-24 r^{2} C_1^{2} \pi  e^{-2B r^{2}} 
+ 8 \pi C_1 r^{2} \rho_{\mathrm{c}} e^{-B r^{2}} 
 - 4 r^{2} \left( 2w + 1 \right)
\sqrt{\rho_{0} C_1} e^{- \frac{1}{2} B r^{2}}
+ \rho_{\mathrm{c}}
\end{aligned}
}
\Bigg]^{1/2}.
\end{align}

\begin{figure*}[t!]
	\centering
	\includegraphics[width=7cm]{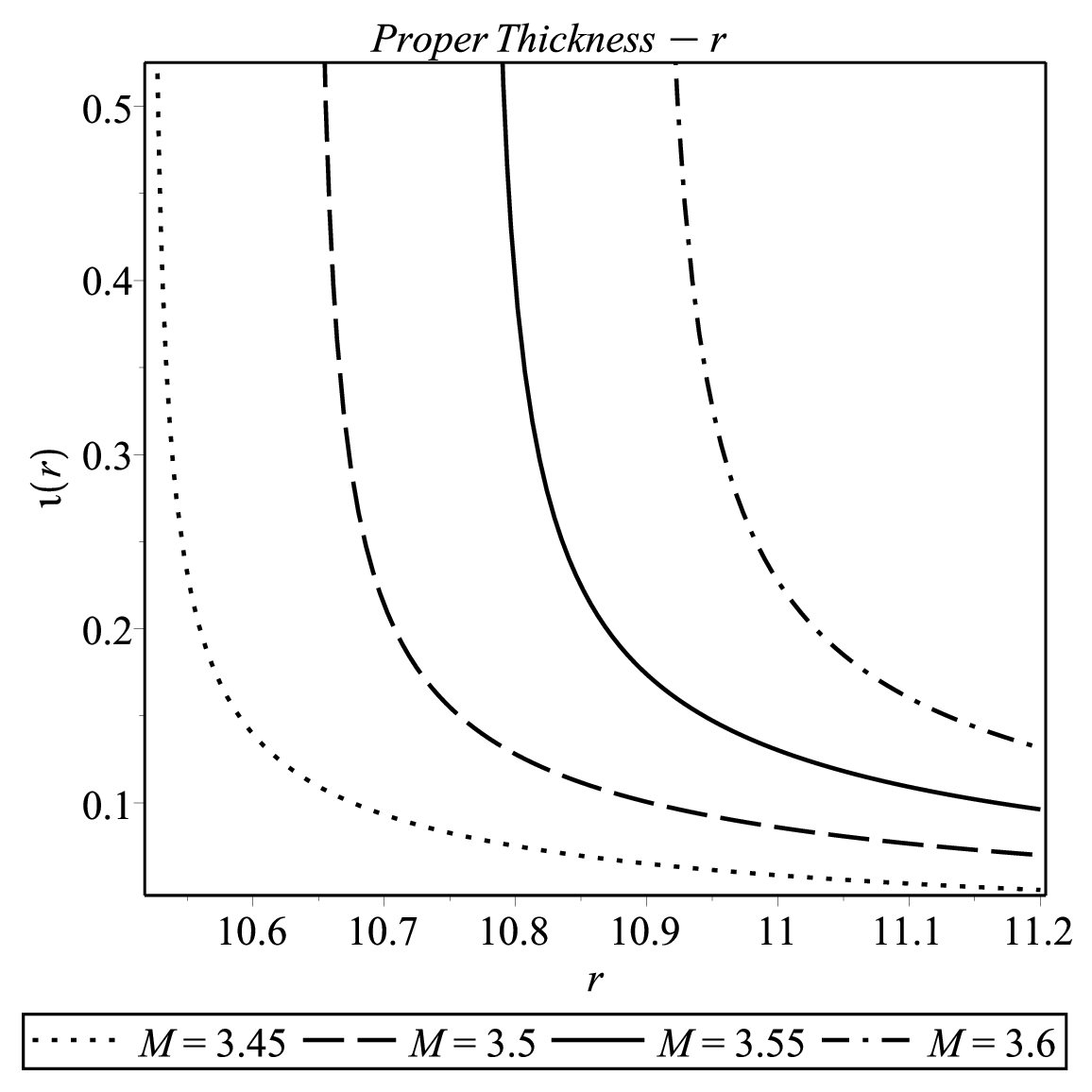}
	\caption{Proper thickness of the gravastar shell, $\ell$, as a function of the radial coordinate $r$ within the shell region $R\leq r\leq R+\epsilon$. The curves represent the numerical evaluation of Eq.~(3.11) for different gravastar masses using the fixed parameters $Q=0.138221\,\mathrm{km}^{2}$ and $\rho_0=0.05\times10^{-7}\,\mathrm{km}^{-6}$ together with the model parameters listed in Table~I. The proper thickness decreases smoothly as a function of $r$ across the shell, indicating that the shell region remains finite and well behaved throughout the gravastar configuration.}
\end{figure*}

The proper thickness quantifies the actual physical width of the gravastar shell as measured by a local observer, rather than a coordinate artifact. A finite, well-behaved thickness ensures that the shell represents a realistic matter layer capable of supporting stresses, storing energy and entropy, and smoothly mediating between the interior and exterior spacetimes. This confirms that the gravastar avoids unphysical surface singularities and provides a stable, physically meaningful alternative to an infinitesimally thin shell. Fig.~3 depicts the behavior of the proper thickness of the gravastar shell
as a function of the radial coordinate within the shell region. The plot shows
that the proper thickness decreases gradually with increasing radial distance,
indicating that the shell becomes progressively thinner toward the outer
boundary. This behavior is consistent with the thin-shell structure of the
gravastar and ensures a smooth transition between the interior de Sitter-like
core and the exterior braneworld geometry.

\section{Boundary and matching conditions} \label{sec4}

The gravastar configuration naturally decomposes into three distinct regions: an interior core, an intermediate shell, and an exterior vacuum spacetime. The shell serves as the transition layer that joins the interior and exterior geometries across a hypersurface. While the spacetime metric is required to remain continuous across this junction, its first derivatives need not be continuous, allowing for the presence of a nonvanishing surface stress–energy tensor localized on the shell.

The appropriate framework for describing such junctions is provided by the Darmois–Israel formalism~\cite{Darmois,Israel}, which relates the intrinsic stress–energy content of the shell to the discontinuity in the extrinsic curvature across the hypersurface. Within this approach, the surface stress–energy arises from the geometric mismatch between the interior and exterior spacetimes and encapsulates the physical properties of the shell.

As one approaches the outer boundary of the gravastar from within the shell, the variation of the energy distribution gives rise to a discontinuity in the extrinsic curvature. This geometric discontinuity manifests itself as an intrinsic surface energy density and surface pressures on the junction hypersurface. The presence of this shell ensures smooth matching of the interior and exterior regions in terms of the metric potentials, although their derivatives may be discontinuous. Consequently, the gravastar remains geodesically complete and is supported by a well-defined matter configuration at the junction at $r=R$.

Following the prescription originally developed by Lanczos~\cite{Lanczos}, the intrinsic surface stress–energy tensor \( S^{i}_{\;j} \) associated with the shell is expressed as
\begin{equation}\label{eq24}
S^{i}_{\;j} = -\frac{1}{8\pi}\left( \kappa^{i}_{\;j} - \delta^{i}_{\;j}\kappa^{k}_{\;k} \right),
\end{equation}
where \( \kappa^{i}_{\;j} \) denotes the jump in extrinsic curvature across the junction. This jump is defined as the difference between the extrinsic curvatures evaluated on the exterior and interior sides of the shell,
\begin{equation}\label{eq39}
\kappa_{ij} = \kappa^{+}_{ij} - \kappa^{-}_{ij}.
\end{equation}

The extrinsic curvature on either side of the shell is computed from
\begin{equation}\label{eq40}
\kappa^{\pm}_{ij}
= - n^{\pm}_{\nu}
\left[
\frac{\partial^{2} X^{\nu}}{\partial \xi^{i} \partial \xi^{j}}
+ \Gamma^{\nu}_{\alpha\beta}
\frac{\partial X^{\alpha}}{\partial \xi^{i}}
\frac{\partial X^{\beta}}{\partial \xi^{j}}
\right]_{S},
\end{equation}

\begin{table}[ht]
\caption{Values of the model parameters. The gravastar mass $M$ is expressed in units of the solar mass ($M_\odot$), the radii $R_1$ and $R_2$ and the shell thickness $\epsilon$ are given in km, while the critical density $\rho_c$, the integration constant $C_1$, and the Kuchowicz parameter $B$ are expressed in units of km$^{-2}$. The parameters $\omega$ and $C$ are dimensionless.}
\centering
\begin{tabular}{ccccccccc}
\hline\hline
$M~(M_\odot)$ &
$R_1~(\mathrm{km})$ &
$R_2~(\mathrm{km})$ &
$\epsilon~(\mathrm{km})$ &
$\omega$ &
$\rho_c~(\mathrm{km}^{-2})$ &
$C$ &
$C_1~(\mathrm{km}^{-2})$ &
$B~(\mathrm{km}^{-2})$ \\
\hline
3.45 & 10.15 & 10.5253 & 0.3753 & 0.48607 & 1.83771 & 0.36272 & 1.51783 & 0.00865 \\
3.50 & 10.15 & 10.6492 & 0.4992 & 0.50132 & 1.93910 & 0.36035 & 1.61956 & 0.00852 \\
3.55 & 10.15 & 10.7800 & 0.6300 & 0.51035 & 2.04107 & 0.35883 & 1.71900 & 0.00836 \\
3.60 & 10.15 & 10.9071 & 0.7571 & 0.52681 & 2.15920 & 0.35688 & 1.83592 & 0.008221 \\
\hline\hline
\end{tabular}
\label{tab:model_parameters}
\end{table}

where \( \xi^{i} \) are the intrinsic coordinates on the junction hypersurface, and \( \Gamma^{\nu}_{\alpha\beta} \) are the Christoffel symbols associated with the spacetime metric.

The unit normal vectors to the shell hypersurface are given by
\begin{equation}\label{eq41}
n^{\pm}_{\nu}
= \pm
\left|
g^{\alpha\beta}
\frac{\partial f}{\partial X^{\alpha}}
\frac{\partial f}{\partial X^{\beta}}
\right|^{-\frac{1}{2}}
\frac{\partial f}{\partial X^{\nu}},
\end{equation}
and satisfy the normalization condition \( n^{\nu} n_{\nu} = 1 \). The shell is defined by the parametric equation \( f(X^{\alpha}(\xi^{i})) = 0 \), with the plus and minus signs corresponding to the exterior and interior spacetimes, respectively.

This formalism enables a consistent and physically transparent determination of the surface energy density and pressures residing on the gravastar shell, thereby completing the matching of the interior and exterior solutions within a fully covariant framework.

The surface energy density of the gravastar on the brane turns out to have the form

\begin{align}\label{eq42}
   \Sigma &= -\frac{1}{4\pi R}\bigg[\sqrt{e^\lambda}\bigg]_-^+ \nonumber\\
   &= -\frac{1}{4\pi R} \Bigg[
       \sqrt{1 - \frac{2M}{R} - \frac{Q}{R^2}}
       - \sqrt{e^{-\lambda(r)}} \Bigg] \nonumber\\
   &= -\frac{1}{4\pi R} \Bigg[
       \sqrt{1 - \frac{2M}{R} - \frac{Q}{R^2}}
       - \sqrt{
         \frac{-4r^{2}\!\left(2w+1\right)\sqrt{\rho_{0}C_{1}}
               + 8\pi C_{1}\!\left(C_{1}-\rho_{\mathrm{c}}\right)r^{2}
               + \rho_{\mathrm{c}}}
              {2B r^{2}\rho_{\mathrm{c}} + \rho_{\mathrm{c}}}
         } \Bigg].
\end{align}

We also compute the surface pressure of the braneworld gravastar which is given as     

\begin{eqnarray}\label{eq43}
\mathcal{P} & =&\frac{1}{16\pi } \bigg[\bigg(\frac{2-\lambda^\prime R}{R}\bigg) \sqrt{e^{-\lambda}}\bigg]_-^+ \nonumber\\
&=& {\frac {R-M}{4\pi R^{2} }{\frac {1}{\sqrt {1-{\frac {2M}{R}}-{
\frac {Q}{{R}^{2}}}}}}}-{\frac {4\sqrt {3} \left( 24{R}^{2}
\sqrt {\rho_{{0}}{C_1}}+16\pi {C_1} \left( {C_1}-\rho
_{{c}} \right) {R}^{2}+3 \rho_\mathrm{c} \right) }{3\pi R  {\sqrt {\rho_{\mathrm{c}} \bigg[{12{r}^{2}\sqrt {\rho_{{0}}{C_1}}+8
\pi {C_1} \left( {C_1}- \rho_\mathrm{c} \right) {r}^{2}+3
 \rho_\mathrm{c} \bigg]}}}}}.
\end{eqnarray}

Following Eq.~(\ref{eq27}), the total mass contained within the thin shell is determined to be

\begin{eqnarray}
\hspace{-5mm}
m_\mathrm{s}&=&4\pi R^2 \Sigma \nonumber\\
&=&{\frac {1}{4\pi R} \left(\sqrt {{\frac {8\pi 
{{C_1}}^{2}{R}^{3}-8\pi {C_1}{R}^{3} \rho_\mathrm{c}+12
\sqrt {\rho_{{0}}{C_1}}{R}^{3}+3 \rho_\mathrm{c}R}{3 \rho_\mathrm{c}R}}}-
\sqrt {1-{\frac {2M}{R}}-{\frac {Q}{{R}^{2}}}} \right)}.
\end{eqnarray}

Table~I summarizes the numerical values of the principal physical parameters
characterizing the gravastar configuration for different choices of the total
mass of the object. For each mass value, the corresponding inner and outer
radii of the shell, $R_{1}$ and $R_{2}$, are obtained from the junction
conditions, from which the shell thickness $\epsilon = R_{2}-R_{1}$ is
determined. It is observed that as the gravastar mass increases, both the inner
and outer radii increase accordingly, indicating that more massive
configurations correspond to larger compact objects. Consequently, the shell
thickness also increases gradually, which is physically reasonable since a
larger gravitational mass requires a wider transition region connecting the
interior de Sitter--like core to the exterior braneworld vacuum geometry.

The table further shows that the critical energy density $\rho_{\mathrm{c}}$ increases
with the mass of the configuration. In the braneworld framework, $\rho_{\mathrm{c}}$ is
directly related to the brane tension and controls the strength of the
high-energy corrections to the effective gravitational field equations.
Larger values of $\rho_{\mathrm{c}}$ correspond to weaker brane corrections, and in the
limit $\rho_{\mathrm{c}} \rightarrow \infty$ the quadratic matter terms vanish,
thereby recovering standard General Relativity. Furthermore, the equation-of-state
parameter $w$ also increases gradually with the mass. Since $w$
characterizes the thermodynamic behavior of the shell matter, its increase
reflects a modification of the pressure--energy density relation required to
sustain equilibrium in more massive gravastar configurations. This behaviour
ensures that the shell matter remains sufficiently stiff to support the
structure against gravitational collapse while maintaining the smooth matching
between the interior and exterior geometries. The integration constants $C$
and $C_{1}$ listed in the table are determined from the matching conditions and
adjust consistently with the other parameters so that the metric functions and
physical quantities remain regular throughout the configuration. These trends
collectively indicate that the obtained parameter values lead to a physically
consistent and stable gravastar structure within the braneworld framework.

This allows us to compute the total mass of the gravastar which can be expressed as 

\begin{align}
M_{\rm Tot} &= \frac{1}{6 \rho_\mathrm{c} R} \Biggl[ 
 8  \pi  R^3  \Sigma  \sqrt{3}  
\sqrt{\rho_\mathrm{c}\left(12  R^2 \sqrt{\rho_0 C_1} + 8  \pi  C_1 (C_1 - \rho_\mathrm{c}) R^2 + 3  \rho_\mathrm{c}\right)}   -12  R^4 \sqrt{\rho_0 C_1} \notag\\
&\quad + \bigg((8  \pi C_1 -48  \pi^2 \Sigma^2 ) R^4 - 3 Q \bigg) \rho_\mathrm{c} 
- 8  \pi  C_1^2  R^4 
\Biggr].
\end{align}

To ensure a smooth and physically consistent gravastar configuration, it is necessary to match the metric functions at the boundaries of the shell. In particular, continuity of the temporal metric component $g_{tt}$ and its radial derivative $\frac{d g_{tt}}{d r}$ at the exterior boundary $r = R_2$ allows us to determine the constants of the model. This yields the conditions
\begin{equation}\label{eq45}
1 - \frac{2 M}{R_2} - \frac{Q}{R_2^2} = C^2  e^{B R_2^2},
\end{equation}
\begin{equation}\label{eq46}
\frac{2 M}{R_2^2} + \frac{2 Q}{R_2^3} = 2 B R_2  C^2  e^{B R_2^2}.
\end{equation}

Similarly, the radial metric component $g_{rr}$ has been matched with the interior metric at $r = R_1$ and with the exterior metric at $r = R_2$, ensuring a smooth interpolation of the geometry across the shell. These matching conditions guarantee that the shell acts as a proper junction between the interior and exterior spacetimes while maintaining geodesic completeness and regularity of the metric. 
For each gravastar configuration, the matching conditions are solved using the corresponding values of the model parameters listed in Table~I together with the fixed parameters $Q = 0.138221\,\mathrm{km}^{2}$ and $\rho_0 = 5.0 \times 10^{-9}\,\mathrm{km}^{-6}$.

Gravastars are primarily motivated as horizonless alternatives to black holes and therefore should be examined in a mass regime relevant to compact objects that would ordinarily be associated with gravitational collapse.
Motivated by this observation, we have performed the numerical analysis and consider the masses $M=3.45,M_{\odot}$, $3.50,M_{\odot}$, $3.55,M_{\odot}$, and $3.60,M_{\odot}$. These values exceed the masses of all securely observed neutron stars and lie beyond the canonical Tolman--Oppenheimer--Volkoff mass range predicted within standard General Relativity. Consequently, such objects are generally expected to be unstable against further gravitational collapse and are more naturally associated with black-hole formation in the absence of additional physics.
The choice of this mass interval is particularly relevant in the present context because the Shtanov--Sahni braneworld model with a timelike extra dimension introduces high-energy quadratic corrections and nonlocal bulk effects that modify the effective gravitational dynamics on the brane. These corrections can significantly alter the final outcome of gravitational collapse relative to General Relativity, thereby allowing the possibility of stable, horizonless compact configurations in a mass range where black-hole formation would otherwise be anticipated. These masses therefore provide a appropriate setting for assessing the viability of braneworld gravastars as black-hole mimickers.
For each mass, the junction conditions have been solved to determine the corresponding values of the model parameters. Consequently, Table~I has been presented and all figures have been generated using the parameter sets obtained from the matching conditions as mentioned in the figure captions.

These constants fully specify the gravastar model, linking the interior de Sitter-like core, the finite shell, and the exterior vacuum with tidal charge. The smooth matching ensures both the continuity of the metric potentials and a physically consistent distribution of matter and stress across the shell, while the derivative matching guarantees that the gravitational field behaves regularly at the outer boundary. The numerically determined values of the model parameters are physically well justified. The relatively small tidal charge indicates that even slight deviations from the Schwarzschild vacuum, induced by bulk gravitational effects projected onto the brane through the Weyl tensor, can lead to noticeable modifications of the braneworld dynamics. Likewise, the small magnitude of the equation-of-state parameter $w$ suggests that the effective fluid representing the bulk contributions on the brane behaves consistently with standard energy conditions and can effectively mimic additional ordinary matter on the brane. 
Furthermore, the critical energy density on the brane, influenced by the brane tension $\sigma$, is found to be small. This highlights that braneworld corrections play a crucial role in shaping the gravastar configuration, placing the system in a regime where standard general relativity is significantly modified. This contrasts with the large $\sigma$ limit, in which general relativity is effectively recovered.

\section{Physical Viability and Acceptability Conditions} \label{sec5}

The physical viability of a compact object like gravastar can generally be assessed using several complementary approaches. In this work, we examine the physical viability of our gravastar model, in which the temporal metric potential is described by the Kuchowicz function, using two distinct criteria. First, we compute the surface redshift of the gravastar and verify that it lies within the range consistent with a stable configuration. Next, we assess the validity of the standard energy conditions throughout the shell and the interior, ensuring that the matter distribution remains physically reasonable. 

\subsection{Surface Redshift}

The surface redshift of the gravastar can be be obtained by using the formula
\begin{eqnarray}\label{eq34}
	Z_{\mathrm{s}}&=&-1+\frac{1}{\sqrt {g_{\it tt}}} =-1+{\frac {1}{\sqrt {{C}^{2}{{e}^{Br^2}}}}}.
\end{eqnarray}

The radial profile of the surface redshift $Z_\mathrm{s}$ from the inner to the outer boundary of the gravastar is illustrated in Fig.~4. For any self-gravitating compact object, it is well known that the surface redshift must satisfy $Z_\mathrm{s} \leq 2$~\cite{Buchdahl1959,Straumann1984,Bohmer2006} in the absence of a cosmological constant. In our analysis within the Shtanov–Sahni braneworld framework, we do not include a $\Lambda$ term, as the effective four-dimensional cosmological constant vanishes due to a fine-tuned cancellation between the negative brane tension and the positive cosmological constant of the five-dimensional de Sitter bulk ($dS_5$).

Using the numerically determined model parameters obtained from the matching and boundary conditions, the surface redshift reaches its maximum at the inner boundary of the gravastar and decreases smoothly across the shell, attaining a finite nonzero value at the outer boundary. The monotonic decrease of $Z_\mathrm{s}$ throughout the shell indicates a well-behaved gravitational potential, while its magnitude remains comfortably below the theoretical upper bound. This behavior not only confirms the regularity of the shell but also provides strong evidence for the physical physical viability of the gravastar configuration.

\subsection{Energy conditions}

As discussed earlier, in the braneworld scenario, the modifications to the gravitational field equations can be interpreted as effective contributions to the matter sector. Consequently, the standard energy conditions must be evaluated with respect to the \emph{effective stress-energy tensor} rather than the original matter source. In this framework, the energy conditions are expressed in Table II.

\begin{figure*}[t!]
	\centering
	\includegraphics[width=7cm]{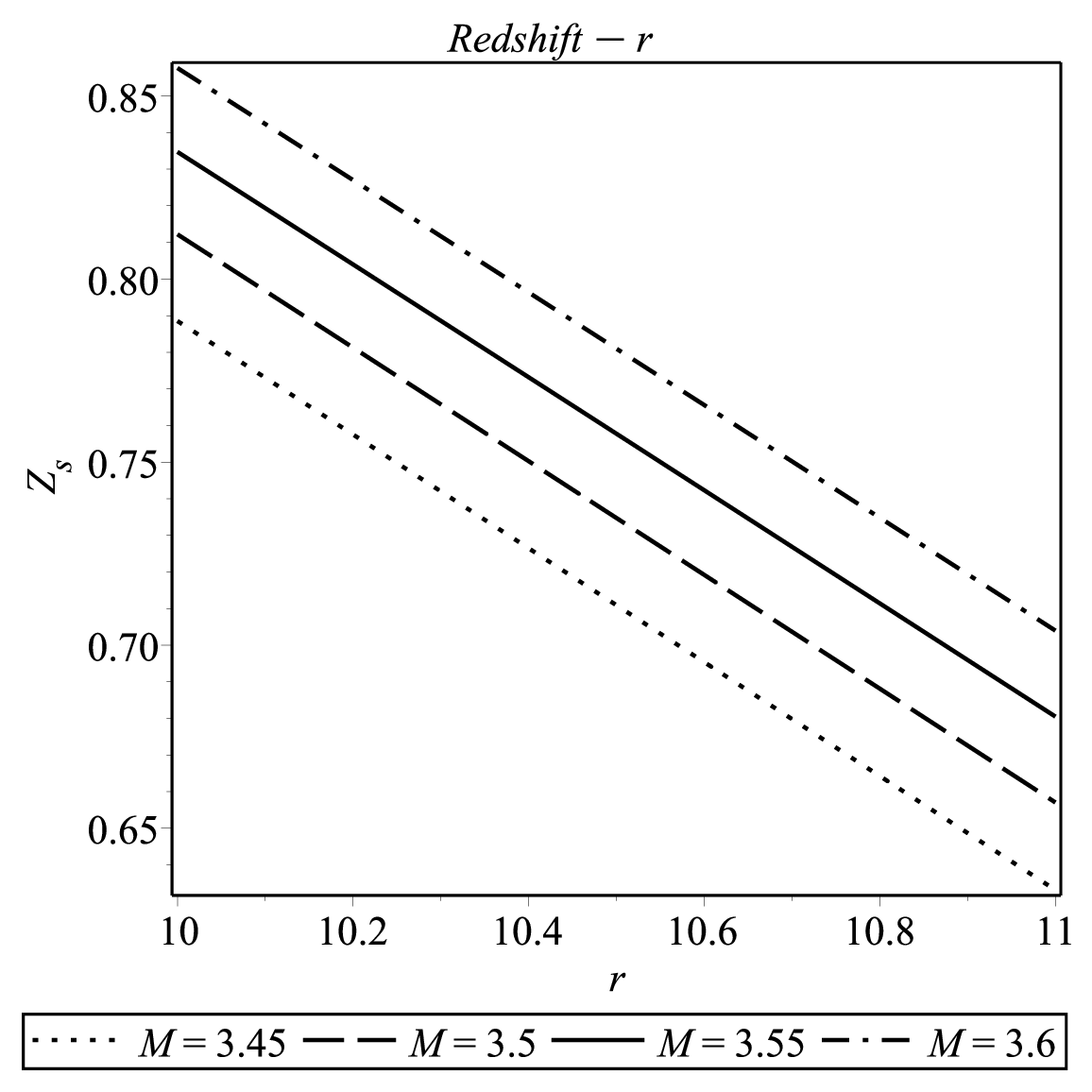}
	\caption{Surface redshift of the gravastar, $Z_s$, plotted as a function of the radial coordinate $r$ across the shell region. The curves represent the numerical evaluation of Eq.~(5.1) for different gravastar masses using the model parameters listed in Table~I. The surface redshift decreases smoothly from the inner boundary of the shell toward the exterior boundary and remains well below the theoretical upper bound $Z_s\leq2$ for compact objects, indicating that the gravastar configuration is physically acceptable and gravitationally stable.}
\end{figure*}

\begin{table}[ht]
\caption{Energy conditions.}
\centering
\begin{tabular}{l@{\hspace{1.5cm}}c}
\hline\hline
\textbf{Energy Condition} & \textbf{Mathematical Form} \\
\hline
Null Energy Condition (NEC) & $\rho^{\mathrm{eff}} \geq 0$ \\
Weak Energy Condition (WEC) & $\rho^{\mathrm{eff}} + p_r^{\mathrm{eff}} \geq 0,\ \rho^{\mathrm{eff}} + p_t^{\mathrm{eff}} \geq 0$ \\
Strong Energy Condition (SEC) & $\rho^{\mathrm{eff}} + p_r^{\mathrm{eff}} + 2p_t^{\mathrm{eff}} \geq 0$ \\
Dominant Energy Condition (DEC) & $\rho^{\mathrm{eff}} - |p_r^{\mathrm{eff}}| \geq 0,\ \rho^{\mathrm{eff}} - |p_t^{\mathrm{eff}}| \geq 0$ \\
\hline\hline
\end{tabular}
\end{table}

Table~II summarizes the standard energy conditions used to assess the physical
viability of the effective matter distribution in the gravastar model.
Since braneworld corrections modify the gravitational field equations,
these conditions are expressed in terms of the effective energy density
$\rho_\mathrm{eff}$ and effective pressures $p_r^\mathrm{eff}$ and
$p_t^\mathrm{eff}$. The table lists the mathematical forms of the
null, weak, strong, and dominant energy conditions used in the analysis.

In Fig.~5, we present the radial profiles of the various linear combinations of the effective energy density and pressures that correspond to these energy conditions across the finite shell. It is evident from the plots that all the physically relevant combinations remain strictly non-negative throughout the shell. This demonstrates that the effective matter supporting the gravastar satisfies all the standard energy conditions, ensuring both the physical plausibility and physical viability of the configuration. 

Physically, this outcome implies that the bulk-induced corrections on the brane do not lead to any exotic or pathological behavior in the shell matter. The effective fluid behaves consistently with ordinary matter in general relativity, providing the necessary pressure and energy density to support the shell while maintaining causal and stable dynamics. The satisfaction of all energy conditions also indicates that the gravastar avoids unphysical stress-energy distributions, such as those that would arise in the presence of negative energy densities or superluminal pressures, reinforcing the viability of the model within a braneworld setting.

\begin{figure}[t!]
\centering
\includegraphics[width=7cm]{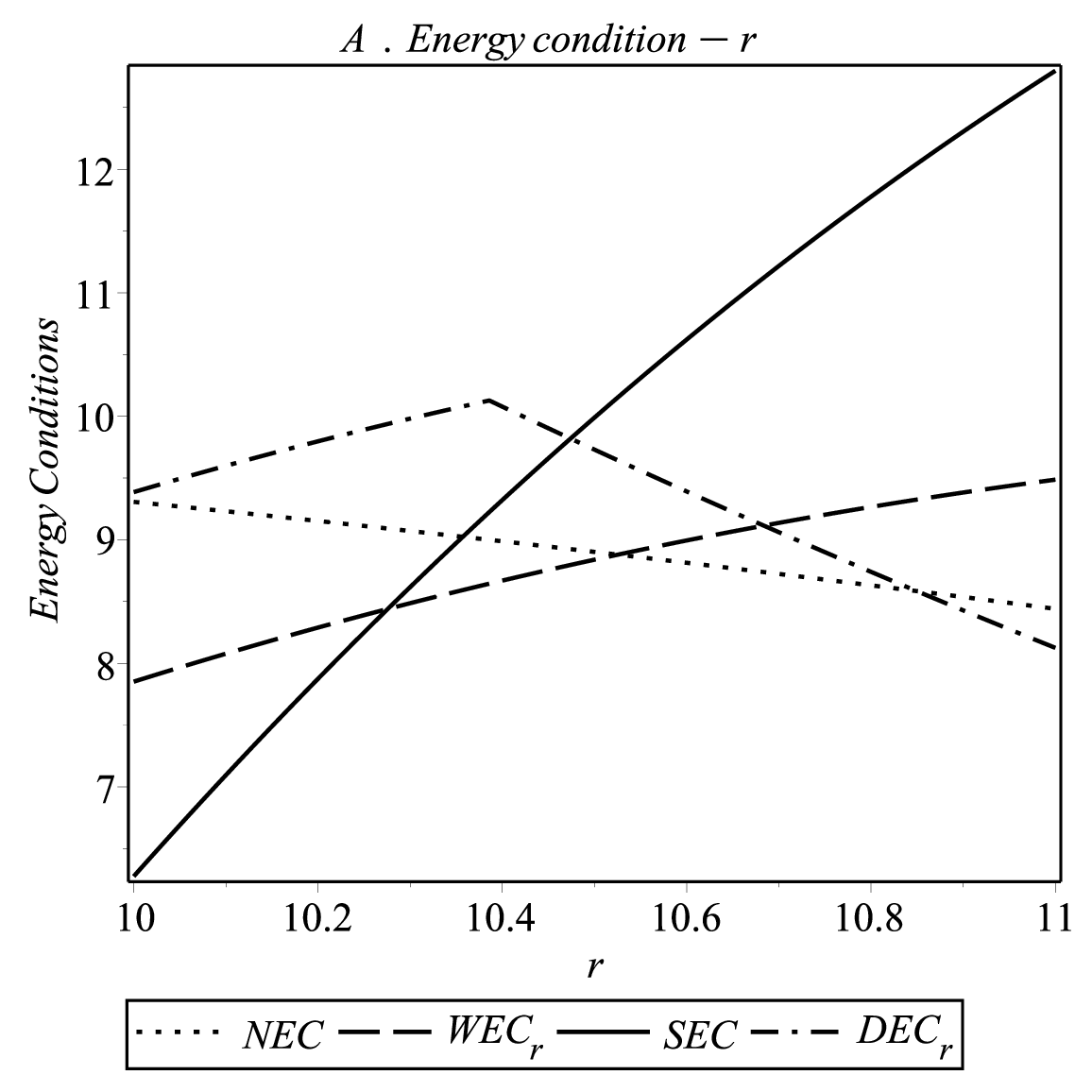}
\includegraphics[width=7cm]{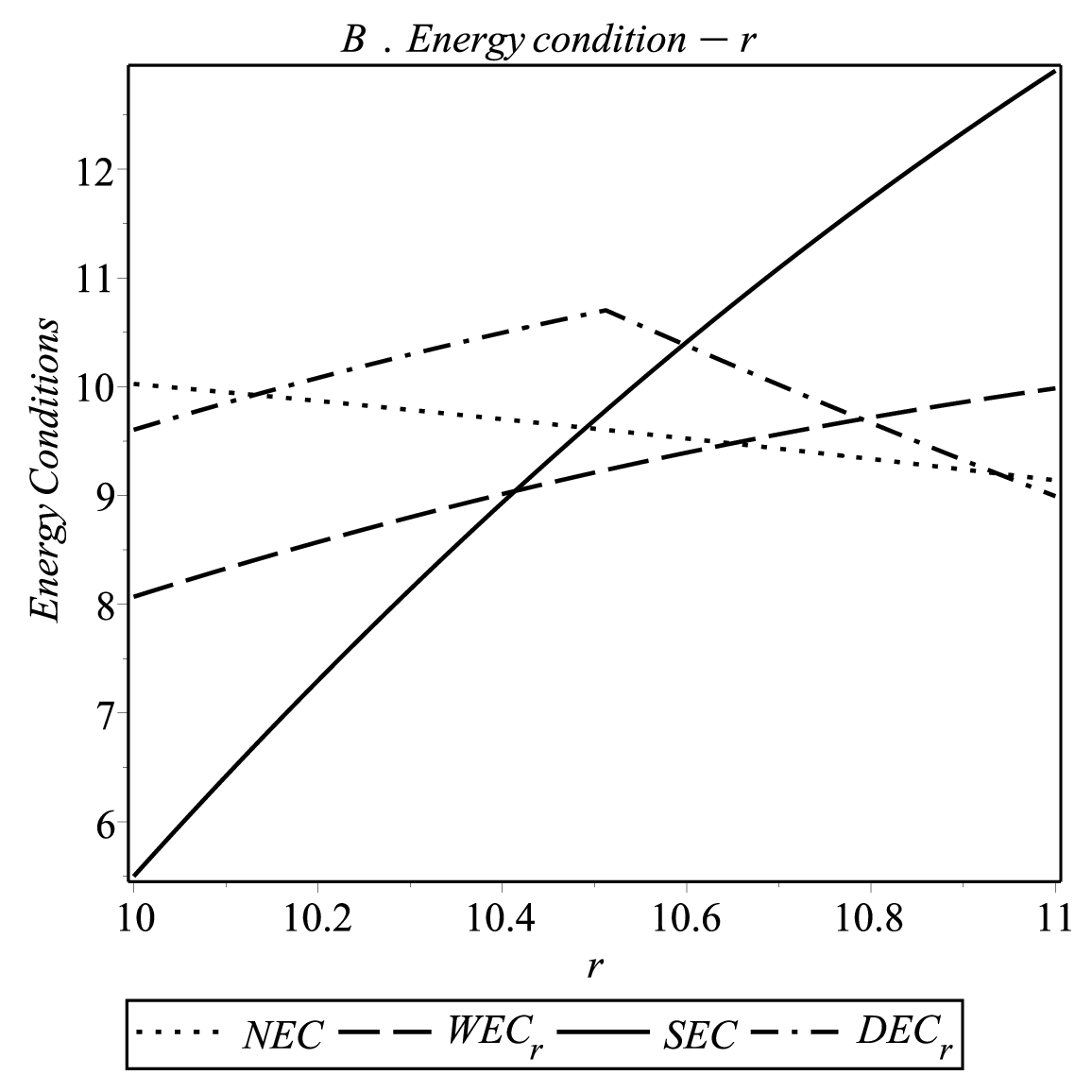}
\includegraphics[width=7cm]{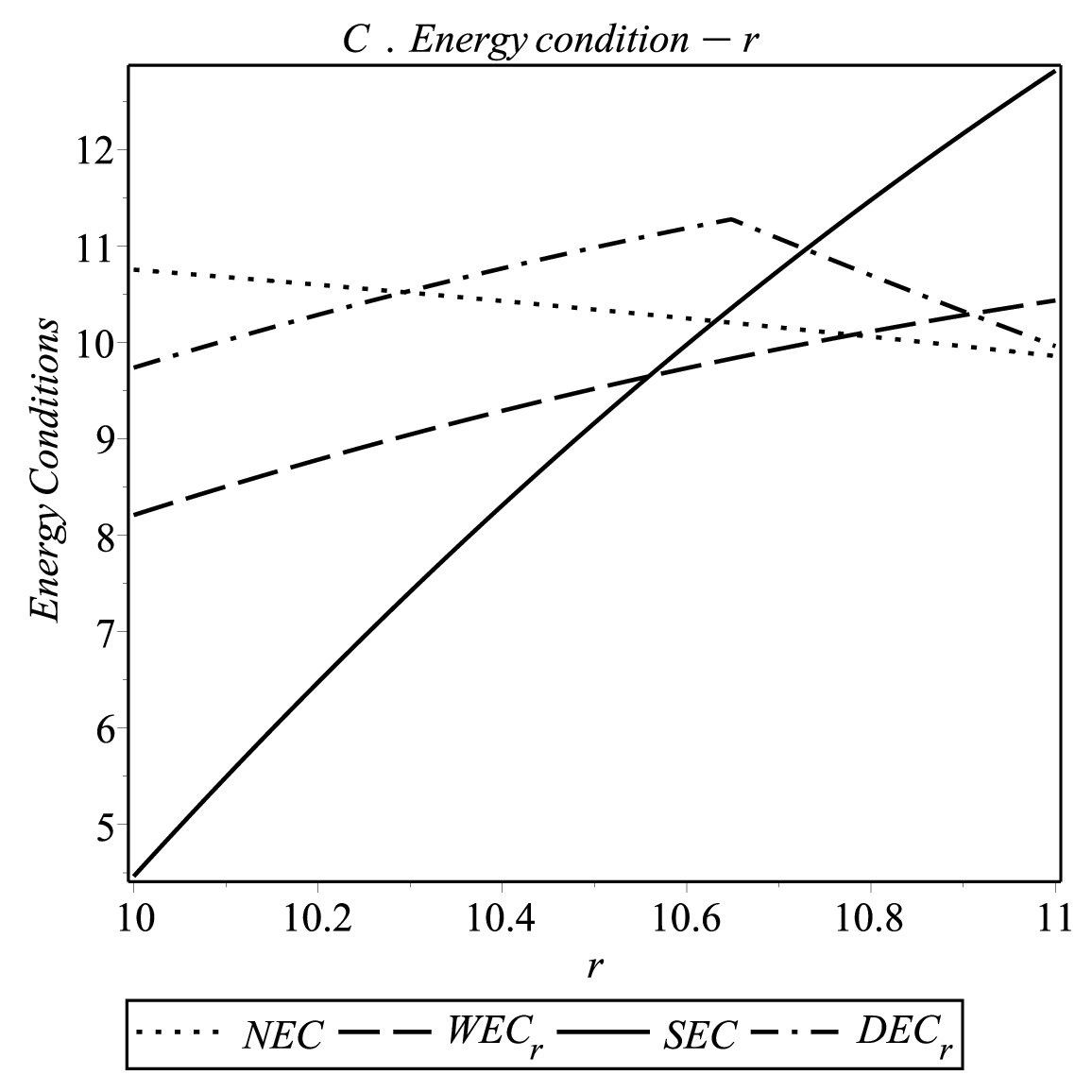}
\includegraphics[width=7cm]{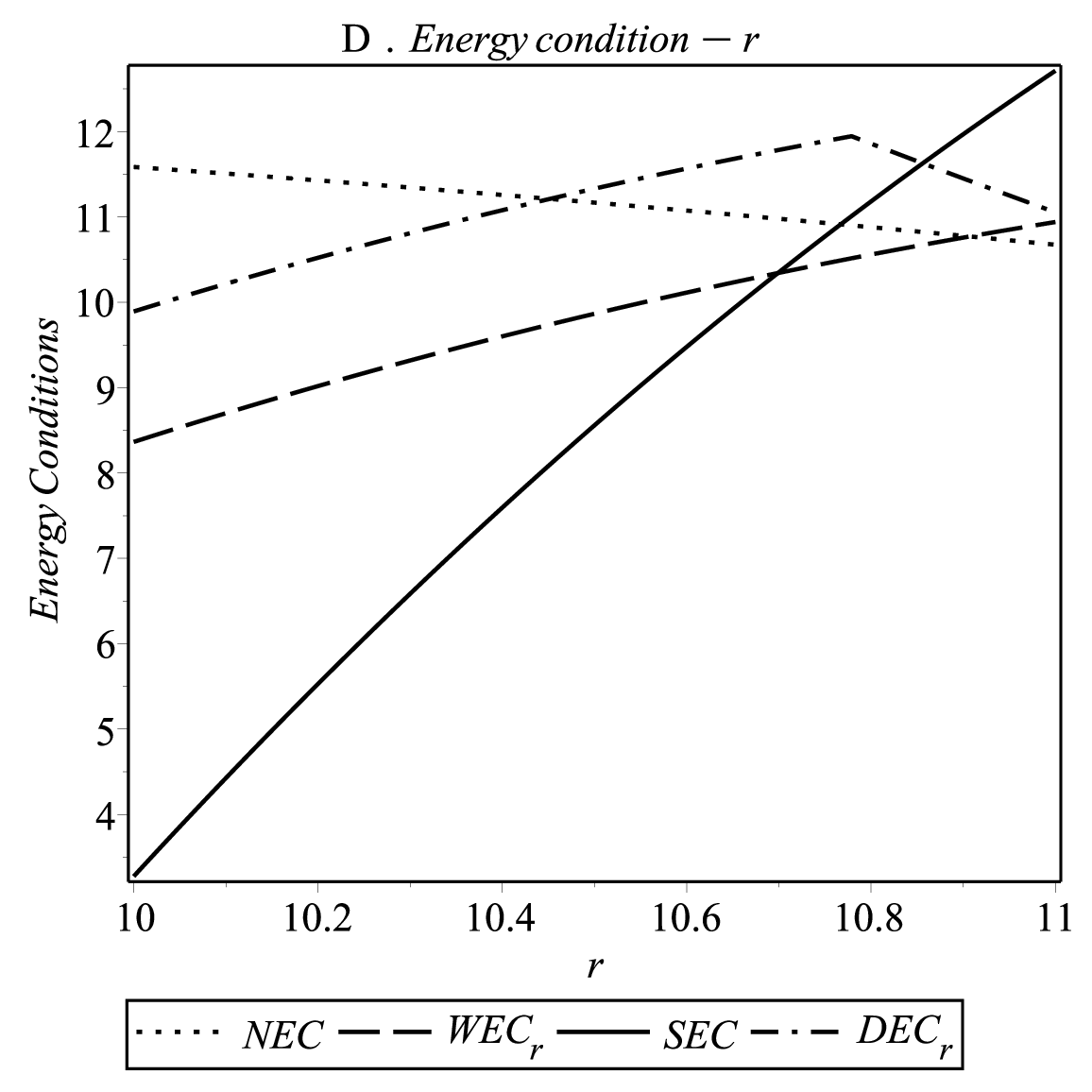}
\caption{Radial variation of the combinations of effective energy density and pressures that define the standard energy conditions as functions of the radial coordinate $r$ within the gravastar shell. The plots are evaluated using the fixed parameters $Q=0.138221\,\mathrm{km}^{2}$ and $\rho_0=0.05\times10^{-7}\,\mathrm{km}^{-6}$ together with the model parameters listed in Table~I. Here, Fig.~5(A) corresponds to $M=3.45\,M_\odot$, Fig.~5(B) to $M=3.50\,M_\odot$, Fig.~5(C) to $M=3.55\,M_\odot$, and Fig.~5(D) to $M=3.60\,M_\odot$.}\label{energy}
\end{figure}

However, these tests do not constitute a genuine stability analysis in the perturbative sense.
The presented results establish the physical viability and consistency of the gravastar solutions rather than their dynamical stability. In particular, the satisfaction of the energy conditions and the bounded behaviour of the surface redshift indicate that the configurations are physically admissible and free from obvious pathologies, but they do not by themselves guarantee stability against perturbations.
A rigorous stability investigation would require a dedicated perturbative treatment, such as a linearized analysis of time-dependent fluctuations of the metric and matter variables, a radial perturbation analysis, or a thin-shell stability study based on the dynamical junction conditions. Such an analysis lies beyond the scope of the present work and will be considered in a future investigation.

\section{Conclusions} \label{sec6}

In this work, we have developed a gravastar model embedded in the SS braneworld, where the timelike nature of the extra dimension fundamentally alters the collapse dynamics. The SS scenario, with its negative brane tension and positive bulk cosmological constant, is distinguished by its ability to generate non-singular bouncing cosmologies. We have shown that this same mechanism, when applied to compact objects, provides a natural resolution of the singularity problem that plagues classical black hole solutions. Motivated by recent work on traversable wormholes in the SS braneworld \cite{Sengupta2023}, which showed that besides cosmological singularities, wormhole singularities can also be resolved in this framework, we demonstrate that gravastars too acquire a natural singularity-free description. 

The original gravastar model proposed by Mazur and Mottola consists of a de Sitter interior, a thin shell of ultra-relativistic matter, and a Schwarzschild exterior within the framework of General Relativity. Subsequently, numerous extensions were developed involving anisotropic fluids, charged gravastars, dark-energy gravastars, and gravastars in alternative theories of gravity. In many of these constructions, pressure anisotropy is introduced phenomenologically at the level of the matter sector in order to satisfy the field equations and maintain equilibrium of the configuration.
In contrast, the present model is formulated within the Shtanov--Sahni braneworld scenario with a timelike extra dimension. A distinctive feature of this framework is that the effective four-dimensional gravitational field equations contain both local quadratic energy-momentum corrections and nonlocal Weyl terms induced by the higher-dimensional bulk geometry. The projected Weyl tensor contributes differently to the radial and tangential sectors of the effective stress-energy tensor, thereby generating an intrinsic pressure anisotropy. Consequently, the anisotropic stresses required for gravastar formation arise naturally from the extra-dimensional gravitational dynamics rather than being imposed \textit{a priori} through a phenomenological anisotropic fluid prescription. This provides a more fundamental geometric origin for the anisotropy that plays a central role in gravastar physics.
A second important distinction concerns the high-energy behaviour of the theory. Owing to the timelike extra dimension, the Shtanov--Sahni model contains negative quadratic energy-density corrections that lead to the existence of a critical density $\rho_c = 2|\sigma|$, which acts as an upper bound on the physically admissible energy density on the brane. Unlike General Relativity, where the matter density may grow without bound during gravitational collapse, the present framework possesses an intrinsic high-energy cutoff determined by the brane tension. Since the effective Ricci curvature is sourced by the effective matter sector, this bounded-density property constrains the growth of the Ricci component of spacetime curvature and naturally favours the avoidance of curvature singularities. The theory therefore provides a particularly suitable arena for studying regular compact objects and horizonless alternatives to black holes. Furthermore, the exterior geometry employed in the present work is the tidal-charged braneworld solution rather than the Schwarzschild spacetime commonly used in conventional gravastar models. The tidal charge parameter encodes bulk gravitational effects and modifies both the matching conditions and the global structure of the resulting compact object. Therefore, compared with the original Mazur--Mottola gravastar, anisotropic gravastar models in General Relativity, and gravastar constructions in other modified-gravity theories, the present framework combines three distinctive ingredients: (i) a geometrically induced pressure anisotropy arising from nonlocal bulk effects, (ii) a fundamental upper bound on the energy density resulting from the timelike extra dimension, and (iii) a tidal-charged braneworld exterior geometry. These features provide additional theoretical motivation for the construction of regular gravastar solutions and constitute the principal novelty of the present work.

Our gravastar construction exhibits several noteworthy features. The interior region is described by a Bose--Einstein condensate with equation of state $p=-\rho$, which contributes negatively to the active gravitational mass and dynamically provides the repulsive behavior required to prevent further gravitational collapse. In contrast to many earlier gravastar models that rely on idealized thin-shell approximations, the present framework admits exact analytic solutions for a shell of finite thickness composed of stiff matter, thereby offering a more realistic description of the transition layer. An additional intrinsic anisotropy arises from the projection of the bulk Weyl tensor onto the brane, leading to anisotropic effective pressures that emerge naturally within the Shiromizu--Maeda--Sasaki braneworld formalism~\cite{SMS} and contribute to enhancing the overall physical viability of the configuration. Furthermore, the model consistently satisfies the required junction and energy conditions, and an analysis of the surface redshift confirms the physical viability and self-consistency of the resulting gravastar solution.

The timelike extra dimension is central to these results. It ensures that the effective cosmological constant on the brane can vanish under fine-tuning, while simultaneously allowing the bulk corrections to regularize the interior geometry. This dual role highlights the SS braneworld as a particularly fertile ground for constructing singularity-free compact objects. Importantly, the induced anisotropy and suppressed effective mass are not ad hoc assumptions but direct consequences of the higher-dimensional corrections. Our findings resonate with recent investigations of traversable wormholes in the SS braneworld \cite{Sengupta2023}, where it was shown that wormhole curvature singularities can be avoided with matter satisfying the energy conditions. Taken together, these results suggest a unifying picture: the SS braneworld provides a consistent framework in which diverse singular structures in the form of cosmological bounces, wormholes, and gravastars can be rendered non-singular. This strengthens the case for the SS model as a promising candidate for probing quantum-gravity inspired phenomena in astrophysics.

We briefly outline a few ideas concerning the possible observational signatures of gravastars, with particular emphasis on the braneworld framework. In scenarios where our four-dimensional universe is embedded in a higher-dimensional bulk spacetime, braneworld gravity has proven useful in interpreting several recent astrophysical observations. Notably, such models have been invoked in the context of the multimessenger event GW170817 observed by the LIGO/Virgo collaboration, along with its electromagnetic counterpart GRB170817A detected by the \textit{Fermi} Gamma-ray Burst Monitor and the \textit{INTEGRAL} Anti-Coincidence Shield spectrometer~\cite{Visinelli}. Furthermore, the observed shadow of the supermassive compact object M87$^{\ast}$ has been shown to admit a consistent explanation within a braneworld scenario featuring an Anti-de Sitter ($AdS_5$) bulk geometry~\cite{Vagnozzi}. These developments motivate the study of gravastars within frameworks that can be regarded as an ultraviolet-corrected extension of general relativity and that have already demonstrated relevance in describing non-singular black hole bounces~\cite{lqc} as well as non-singular bouncing cosmological models~\cite{Sahni4}. In this context, gravastars emerge as compelling horizonless alternatives to classical black holes, potentially compatible with quantum gravity–inspired modifications at high curvature scales.

Although no direct observational evidence for gravastars currently exists, several indirect detection strategies have been proposed in the literature. These include the analysis of gravastar shadows~\cite{Sakai2014}, gravitational microlensing signatures exhibiting a higher maximal luminosity than those produced by black holes of identical mass~\cite{Kubo2016}, and characteristic features in gravitational-wave ringdown signals. In particular, it has been suggested that the ringdown phase of the GW150914 event~\cite{Abbott2016} could, in principle, be generated by a horizonless compact object such as a gravastar~\cite{Cardoso1,Cardoso2}, although this interpretation remains under active debate~\cite{Chirenti2016}. Taken together, these studies indicate that while gravastars remain observationally elusive, current and future high-precision gravitational-wave and electromagnetic observations may provide viable avenues for distinguishing them from classical black holes, especially within modified gravity frameworks such as the braneworld scenario considered in the present work.

In conclusion, the gravastar on the SS brane offers a compelling, physically consistent alternative to black holes, rooted in a framework that naturally resolves singularities. Our analysis reveals, in a fully analytic and self-consistent manner, how non-local bulk gravitational effects encoded through the projected Weyl tensor can dynamically generate intrinsic pressure anisotropy inside the gravastar, without the need for ad hoc anisotropic matter sources. This anisotropy, together with the repulsive contribution of the Bose--Einstein condensate core characterized by $p=-\rho$, plays a central gravitational role by weakening the effective attractive pull and enhancing the physical viability of the configuration against collapse. Unlike previous gravastar models, where physical viability is often imposed through phenomenological assumptions or fine-tuned junction conditions, here it emerges naturally from the modified gravitational sector itself, providing a genuinely geometric origin for the stabilizing mechanism.  The existence of an exact finite-thickness stiff-matter shell further distinguishes our construction from past works relying on infinitesimally thin shells, allowing a more realistic description of intermediate regions and clarifying the physical interpretation of surface stresses, redshift bounds, and energy conditions in a higher-dimensional setting. 

From a gravitational physics perspective, our results demonstrate that braneworld corrections can effectively mimic exotic matter behavior while remaining consistent with standard energy conditions when interpreted through the effective stress--energy tensor, thereby offering a new avenue to reconcile compact object models with fundamental gravitational theory. Its distinctive features like repulsive interior, finite-thickness shell, and anisotropy induced by bulk effects make it an attractive candidate for further theoretical and observational investigation, and a strong contender in the ongoing search for singularity-free compact objects in high-energy gravity. Looking ahead, this framework opens several promising directions, including dynamical physical viability analyses under radial and non-radial perturbations, potential observational imprints distinguishing braneworld gravastars from classical black holes through lensing or ringdown signatures, and extensions to rotating or charged configurations. Gravitational wave astronomy and precision astrophysical probes may provide the necessary discriminants to distinguish SS gravastars from classical black holes, opening a new observational window into quantum-gravity motivated compact objects. More broadly, our findings suggest that gravastars may serve as sensitive probes of extra-dimensional gravity, providing a novel bridge between high-energy gravitational physics and astrophysical phenomenology.

\section*{Acknowledgement}

RS dedicates this work to the memory of his beloved grandmother Kana Dutta who passed away on December 30, 2025. RS thanks the Council for Scientific and Industrial Research (CSIR), Government of India for financial support through the Research Associateship scheme. The work of KB was supported in part by the JSPS KAKENHI Grants No. 24KF0100 and No. 25KF0176,  Competitive Research Funds for Fukushima University Faculty (25RK011), and a grant-in-aid of academic research of Yamaguchi Scholarship Foundation. SG is thankful to the Directorate of Legal Metrology under the Department of Consumer Affairs, West Bengal for their support.

\end{document}